\documentclass[prd,twocolumn,superscriptaddress,floatfix,showpacs,nofootinbib]{revtex4-2}
\usepackage{amsfonts}
\usepackage{amsmath}
\usepackage{amssymb}
\usepackage{xcolor}
\usepackage{graphicx}
\usepackage[normalem]{ulem}
\font\mybb=msbm10 at 10pt
\def\bb#1{\hbox{\mybb#1}}
\def\be{\begin{equation}}
\def\ee{\end{equation}}
\def\bea{\begin{eqnarray}}
\def\eea{\end{eqnarray}}

\begin{document}
\title{Tensionless  M5-brane in light-cone gauge. \\  {\small  The spinor moving frame method is wanted}}

\author{Igor Bandos}
\email{igor.bandos@ehu.eus}
\affiliation{Department of Physics and EHU Quantum Center, University of the Basque Country UPV/EHU,
P.O. Box 644, 48080 Bilbao, Spain,}
\affiliation{IKERBASQUE, Basque Foundation for Science,
48011, Bilbao, Spain }

\date{{\small November 23, 2026}}


\begin{abstract}
We perform the complete light--cone gauge fixing  for the tensionless limit of the M-theory 5-brane the action for which was recently obtained in \cite{Bandos:2026pdg}. The problem of gauge fixing of local fermionic $\kappa$--symmetry becomes quite nontrivial because the  projector
of that is constructed from the gauge field strength which is required to be nonvanishing and nontrivial in this limit for its consistency. 
To resolve  this problem the spinor moving frame method has been adapted.
The corresponding moving frame is constructed on a vector quadratic in the gauge field strength, which is  required to be  nonvanishing and nondegenerate in the tensionless limit, and the derivative of the PST (Pasti-Sorokin-Tonin) scalar. The symmetries of the gauge fixed action and equations of motion are briefly discussed.

To make a comparison with the case of usual tensionless p-branes (null-p-branes), those  which do not carry worldvolume gauge fields, we also elaborate their spinor moving frame formulation.
\end{abstract}

\maketitle

\thispagestyle{empty}




\section{Introduction}

The tensionless limit of M5--brane was the subject of interest almost three decades, beginning  \cite{Gibbons:2000ck} where the starting point was the superembedding description of M5-brane \cite{Howe:1996yn,Howe:1997fb}.
There this limit was taken in the simplest case of frozen  scalar and fermionic fluctuations of the M5-brane i.e. when  the theory is reduced to chiral 6d conformal electrodynamics. In a more  recent  \cite{Townsend:2019ils} the tensionless limit for this case of a planar and static 5-brane was studied in the frame of the Hamiltonian approach \cite{Bergshoeff:1998vx}.

In the very recent \cite{Bandos:2026pdg} we obtained a complete supersymmetric tensionless M5-brane action by taking a zero-tension limit of the original super-M5-brane action  \cite{Bandos:1997ui,Aganagic:1997zq}. The consistency of the procedure requires the gauge field strength of the 2-form worldvolume gauge field to be nonvanishing and in certain sense  nondegenerate.
This implies that tensionless limit can be also considered as strong field limit of the M5--brane.
The tensionless action possesses spacetime supersymmetry and $\kappa$--symmetry, which guaranties that the ground state  of tensionless M5--brane is a stable 1/2 BPS state preserving a half of supersymmetry. We have also shown in  \cite{Bandos:2026pdg} that the conformal invariance of the chiral 6d electrodynamics observed in \cite{Gibbons:2000ck} and  \cite{Townsend:2019ils}, survives when the bosonic fluctuations of the tensionless M5--brane (but not the fermionic ones) are alive, although in a deformed form as the conformal transformations of $6d$ worldvolume coordinates become field dependent. To this end we had to gauge fix the diffeomorphism symmetry of the bosonic tensionless M5--brane.

To perform a similar study in a complete tensionless M5--brane model, including the fluctuation of worldvolume fermionic fields, one needs to perform gauge fixing of the local fermionic $\kappa$--symmetry. This happen to be   quite challenging problem and was not addressed in \cite{Bandos:2026pdg}. The main aim of this paper is to resolve this issue and develop a complete gauge fixing description of the tensionless M5--brane.

The light-cone gauge is  widely and effectively used in the brane physics. However, in the case of tensionless M5-brane \cite{Bandos:2026pdg} its fixing is not straightforward because, as we have already noted, the structure of the $\kappa$--symmetry projector in this case involves essentially the worldvolume gauge field strength   \cite{Bandos:2026pdg}. This implies that the gauge fixing of light-cone gauge on fermionic coordinates has to be made differently for different ranges of values of the gauge field strength.

This peculiarity, in particular, does not allow the straightforward use of the light-cone gauge description of M5-brane developed in \cite{Bandos:2008fr}; actually it is reasonable to expect that the tensionless limit and the light-cone gauge fixing procedures do not commute.

To resolve this issue and to perform the light-cone gauge fixing of the  $\kappa$--symmetry of the tensionless M5--brane, we have developed a version of the spinor moving frame approach of \cite{Bandos:1992np,Bandos:2007wm,Bandos:2014lja}
appropriate for this system. It is based on moving frame build on two unit vectors, one constucted from  the derivative of the PST (Pasti-Sorokin-Tonin) scalar, $v^{\underline{a}}=v^m E_m^{\underline{a}}$,
which we chose to be timelike in \cite{Bandos:2026pdg} and in this paper, and the second, $\hat{s}^{\underline{a}}={s}^m E_m^{\underline{a}}/\sqrt{-s^ns_n}$, where
$s_m$ is constructed from the bilinear of the field strengths of the worldvolume gauge field (see \eqref{sm:=} below).
The consistency of the  tensionless limit of M5--brane requires this vector to be nonvanishing, $s_m \not=0$, and hence, the gauge field strength to be nonvanishing and nondegenerate  \cite{Bandos:2026pdg}.

We also use the above discussed moving frame to fix an appropriate gauge with respect to the worldvolume diffeomorphism symmetry. The light-cone gauge fixed form of the equations of motion and the residual symmetries preserving the gauge fixing conditions are also discussed.

\bigskip

To compare the tensionless M5-brane  with the usual tensionless p-branes  (null-p-branes) which do not carry worldvolume gauge fields. To this end, we elaborate their spinor moving frame formulations,  analyze their symmetries and discuss gauge fixing. In particular, we show that the induced metric of null-p-brane is degenerate in this formulation, like it is the ILST (Isberg-Lindstr\"{o}m-Sundborg-Theodoridis) description \cite{Isberg:1993av,Lindstrom:2026eyz}, and in contrast to the case of tensionless M5-brane \cite{Bandos:2026pdg} in which the induced metric remains nondegenerate. The light-cone gauge fixing conditions for the coordinate functions of null-p-branes and the residual symmetry acting on these are shown to be essentially the same as in the case of tensionless M5-brane.


\begin{widetext} 
\section{Tensionless M5-brane action}

The complete supersymmetric action for the tensionless limit of M5--brane obtained in  \cite{Bandos:2026pdg} reads

\begin{eqnarray} \label{SM5-T0=}
S_{M5} \vert_{T\mapsto 0} &=&  -  \int_{W^6} d^6\xi \sqrt{-g}\sqrt{-s^ms_m}
-
\frac 1 2 \int_{W^6} \, v \wedge H_3 \wedge i_vH_3   \qquad \nonumber \\
 &=&  -  \int_{W^6} d^6\xi \sqrt{-g}\sqrt{-s^ms_m}
-
\frac 1 4
\int_{W^6} \, d^6\xi \sqrt{-g}\tilde{H}^{mn} H_{mnp}v^p  \; . \qquad
\end{eqnarray}
In it $\xi^m$ ($m=0,1,2,3,4,5$) are worldvolume coordinates of the $6d$ worldvolume $W^6$, the parametric  description of which as a surface in 11D superspace with coordinates $Z^M=(X^{\underline{m}}, \Theta^{\underline{\alpha}})$ is given in terms of coordinate functions
\be
Z^M(\xi) =(X^{\underline{m}}(\xi), \Theta^{\check{\underline{\alpha}}}(\xi)) \; , \qquad
{\underline{m}}=0,1,...,9,10\, \qquad \check{\underline{\alpha}}=1,..., 32 \; . \qquad
\ee
These coordinate functions enter the action  \eqref{SM5-T0=} through the determinant $g=\det (g_{mn})$ of the induced metric
\begin{eqnarray}
\label{g=ind}
g_{mn}(\xi)= {E}_m^{\underline a} \eta_{\underline{ab}}{E}_{n}^{\underline b} ={E}_m^{\underline a} {E}_{n\underline a}.
\end{eqnarray}
constructed from the pull--back $E^{\underline a}({Z}(\xi))=d\xi^m {E}^{\underline a}_m$  of the bosonic supervielbein one-form $E^{\underline a} (Z)= dZ^M E_M^{\underline a}(Z)$ of 11D supergravity, ${\underline a}=0,1,\ldots, 9$. Notice that the pull--backs of the 3-form and 6--form gauge fields of the 11D supergravity, $A_3$ and $A_6$, which are present in the Wess-Zumino term of the M5--brane action \cite{Bandos:1997ui} disappear in the tensionless limit.

In this paper we will be interested in the case of tensionless M5 in flat 11D superspace  in which  $E^{\underline a}$ and its fermionic counterpart $E^{\underline{\alpha}}$ can be identified with 11D version of Volkov-Akulov 1-form and with derivative of the fermionic coordinate, respectively,
\begin{eqnarray}
\label{Ea=}
E^{\underline a} =dX^{\underline m}\, \delta_{\underline m}{}^{\underline a} - i
d\Theta^{\underline\alpha} \Gamma^{\underline a}_{\underline\alpha\underline\beta}\Theta^{\underline\beta}\; , \qquad E^{\underline\alpha}=d\theta^{\underline\alpha}\; , \qquad {\underline\alpha}=1,\ldots ,32\; .
\end{eqnarray}
Here we used the real symmetric  $\Gamma^{\underline a}_{\underline\alpha\underline\beta}$ which are constructed from $11D$ gamma-matrices in the Majorana representation and charge conjugation matrix as described in Appendix \ref{app=11DGamma}.
\end{widetext}

$H_3$ in  \eqref{SM5-T0=} is the field strength of the worldvolume 2-form gauge potential
$b_2=  \frac 1 {2} d\xi^n\wedge d\xi^m b_{mn}$,
\begin{eqnarray}
\label{H3} H_3 &=& db_2 = \frac 1 {3!}d\xi^p\wedge d\xi^n\wedge d\xi^m H_{mnp} \;  \qquad
\end{eqnarray}
and the worldvolume 1-form  \begin{eqnarray}\label{v:=} v &= &
d\xi^m v_m = \frac {da(\xi)} {\sqrt{ \partial_n a \, g^{nl}\, \partial_l a}}\; , \qquad v_mv^m=1
\qquad
\end{eqnarray}
is constructed from the derivative of the PST scalar $a(\xi)$,  an auxiliary scalar field  which is subject to the 'topological' restriction  $\partial_m a(\xi)\partial^m a(\xi)\not =0$ \footnote{The choice of unit vector $v_m$  to be time-like is not mandatory: spacelike choice is equally possible \cite{Perry:1996mk,Pasti:1997gx,Bandos:1997ui,Aganagic:1997zq} in topologically admissible cases \cite{Bandos:2014bva}.}. The inverse of the induced metric \eqref{g=ind} is denoted by $g^{mn}(\xi)$ and  used to raise the worldvolume vector indices, e.g. $v^m= g^{mn}v_m$.

In the second line of \eqref{SM5-T0=}
\begin{eqnarray}
\label{tH=}
\tilde{H}^{mn} & = & *H^{mnr} v_r = \frac 1 {3!\sqrt{-g}}
\epsilon^{mnrlpq} H_{lpq}v_r \; , \qquad \end{eqnarray}
which is also involved in the definition of the six--vector
\begin{equation}\label{sm:=}
s_m \equiv \frac {\sqrt{-g}}{8}
\epsilon_{mnklpq} \tilde{H}^{nk}\tilde{H}^{lp}v^q\;
\end{equation}
which enters the  first term of the action \eqref{SM5-T0=}. By construction, this vector is orthogonal to $v_m$,
\begin{equation}\label{tv=0}
s_m v^m=0\; ,
\end{equation}
and hence, space-like for our choice of timelike $v$.

By construction,  the tensionless M5--bane action is also invariant under spacetime super-Poincare transformations. It is also not surprising that it is invariant under scaling symmetry \cite{Bandos:2026pdg}  so that the manifest rigid symmetry transformation leaving invariant \eqref{SM5-T0=} is given by
\bea
\label{vX=sPoin+scal}
\delta X^{\underline{a}}= a^{\underline{a}} +  X^{\underline{b}}
L_{\underline{b}}{}^{\underline{a}}+\lambda X^{\underline{a}}+ i \Theta\Gamma^{\underline{a}}\epsilon\, , \qquad
\\
\label{vTh=sPoin+scal}
\delta \Theta^{\underline{\alpha}}= \epsilon^{\underline{\alpha}} + \frac 1 2 \lambda  \Theta^{\underline{\alpha}}-\frac 1 4 L^{\underline{a}\underline{b}}\,  \Theta^{\underline{\beta}}
\Gamma_{\underline{a}\underline{b}}{}_{\underline{\beta}}{}^{\underline{\alpha}}\,
, \qquad
\eea
where $L^{\underline{a}\underline{b}}=-L^{\underline{b}\underline{a}}$ parametrize the Lorentz group,  while $a^{\underline{a}}$ and $ \lambda $ are translation and scaling parameters.

As far as the worldvolume {\it gauge} symmetry transformations are concerned,
besides being manifestly invariant under the  worldvolume diffeomorphisms, the action \eqref{SM5-T0=} possesses local fermionic $\kappa$--symmetry the transformation of which is  given by fermionic variations of coordinate functions,
\begin{eqnarray}\label{kappa=def}
&& i_\kappa E^a := \delta_kZ^M  E_M^a =0  \; , \qquad \nonumber
\\ && i_\kappa E^\alpha := \delta_kZ^M E^\alpha =: \kappa^\alpha \; , \qquad
\end{eqnarray}
the fermionic paremeter of which obeys
\bea && \kappa^\alpha = - \tilde{\mathbf{\Gamma}}^{\alpha}{}_{\beta}  \kappa^\beta \; \label{kap=-bGkap}
\eea
with
\begin{eqnarray}\label{Gkappa-T0=}
 \tilde{\mathbf{\Gamma}}&=&
- \frac {1} {\sqrt{-s^ls_l}} s_{m}\tilde{\Gamma}{}^{mn}v_n \equiv - \frac {1} {\sqrt{-s^ls_l}} (s\tilde{\Gamma}) (v\Gamma) \;, \qquad \nonumber \\ && \tilde{\mathbf{\Gamma}}^{2}={\bb I}_{32\times 32}\,, \qquad 
\end{eqnarray}
where ${\bb I}_{32\times 32}$ denotes unity matrix $\delta^{\underline\alpha}{}_{\underline\beta}$.
The worldvolume gauge field $b_2$ of tensionless M5--brane is inert under this $\kappa$--symmetry,
\be \label{kappa=def-b} \delta_\kappa b_2= 0 \; . \qquad \ee

Note  that  $\sqrt{-s^ls_l}$ appears in the denominator of the $\kappa$--symmetry projector, and also in the  denominators of the Lagrangian equations of motion. This  suggests that in general
\be\label{t2=not0}
\sqrt{-s^ls_l}\not =0\,
\ee
which implies that  the $6\times 6$ antisymmetric matrix $\tilde{H}^{mn}$ (\ref{tH=}), by construction obeying $\tilde{H}^{mn}v_n=0$, should be of maximal rank 4. The tensionless M5--brane theory described  by the action \eqref{SM5-T0=} is consistent when the  gauge field strength is nonvanishing and,  moreover, not degenerate in the  sense of \eqref{t2=not0}. So will be the consistency conditions for our constructions and results below.

The action \eqref{SM5-T0=} is also invariant under two  PST gauge symmetries \cite{Pasti:1995ii,Pasti:1996vs}. The first PST symmetry acts on the fields with a scalar parameter $\varphi(\xi)$
\begin{eqnarray}\label{PST1==}
\delta_\varphi b_2=  \varphi(\xi)\;  \frac {(i_vH_3+{\cal V}_2)}{\sqrt{\partial a\partial a}}  \; , \qquad \delta_\varphi a(\xi)=\varphi(\xi) \; , \qquad \\ \nonumber \delta_\varphi Z^M(\xi)=0\;  ,\qquad
\end{eqnarray}
where $i_v H_3  = \frac 1 {2}d\xi^p\wedge d\xi^n\; v^m H_{mnp} \;  $ and  $
 {\cal V}_2 =\frac 1 2 d\xi^n \wedge d\xi^m  {\cal V}_{mn}$
with \begin{equation}\label{cV2=}  {\cal V}_{mn}=- \frac {\partial \sqrt{-s^ls_l}} {\partial \tilde{H}^{mn}}= \; \frac {\sqrt{-g}}{ 2\sqrt{-s^ss_s} } \epsilon_{mnrlpq} s^{r}v^l \tilde{H}^{pq}\;.
\end{equation}
It makes PST scalar $a(\xi)$ to be pure gauge or Stueckelberg field (which can be gauge fixed to  $a(\xi)=\xi^0$, but not to constant or to a coordinate in a light-cone direction).
The second PST symmetry has a
 one-form parameter $\phi_1=d\xi^m\phi_m(\xi)$ and acts on the 2-form gauge field only,
\begin{equation}
\label{PST2==}
\delta_{\phi_1} b_2= da\wedge \phi_1\; , \qquad \delta_{\phi_1} a(\xi)=0\; , \qquad \delta_{\phi_1} Z^M(\xi)=0\; .  
\end{equation}

The equations of motion of the bosonic and fermionic coordinate functions $X^{\underline {m}}(\xi)$, $\Theta^{\underline{\alpha}}(\xi)$ are
\begin{eqnarray}
\label{bEq=}
\partial_m (T^{mn}{E}_{n\underline{a}})=0\; , \qquad
\\ \label{fEq=}
({E}_{m}\Gamma^{\underline{a}})_{\underline{\alpha}}  { T}^{mn}{E}_{n\underline{a}}=0\; , \qquad
\end{eqnarray}
where the energy--momentum tensor of tensionless M5--brane  $T^{mn}$  has the following form
\begin{eqnarray}
\label{cTmn=}
{T}^{mn}
&=&  -\, \sqrt{-g}\,\sqrt{-s^ls_l}\,(v^m -\, \hat s^m)(v^n  -\, \hat s^n)\;,\qquad \\ &&
\label{hsm:=}\hat{s}^m=s^m/ \sqrt{-s^ls_l}\,.
\end{eqnarray}

The Lagrangian equation of motion of $b_2$ is
\begin{equation}\label{eomb2}
 da \wedge d \left(\frac {i_vH_3+{\cal V}_2}{\sqrt{\partial a\partial a}}\right)=0.
\end{equation}
Formally integrating it and using  the gauge symmetry (\ref{PST2==}) one reduces it to the nonlinear generalization of the self-duality  condition
(see \cite{Bandos:2026pdg} and refs therein):
\begin{eqnarray}\label{ivH3=-cV2}
i_vH_3+{\cal V}_2=0 \qquad \Leftrightarrow \qquad H_{mnk}v^k=-{\cal V}_{mn}\; ,
\end{eqnarray}
with ${\cal V}_{mn}$ from Eq.  (\ref{cV2=}).

\section{Gauge fixing  kappa--symmetry and worldvolume diffeomorphisms}

\subsection{Factorizing  kappa--symmetry projector} \label{fac-kappa}

The fermionic coordinate function transforms under
supersymmetry and $\kappa$--symmetry  \eqref{kappa=def}
as follows
\be\label{vTh=ep+kap}
\delta \Theta^{\underline{\alpha}} (\xi) = \epsilon^{\underline{\alpha}}+ \kappa^{\underline{\alpha}}(\xi)\; ,
\ee
where the parameter of the  local fermionic $\kappa$--symmetry obeys \eqref{kap=-bGkap}. Let us write this condition as
\be\label{kappa==} \kappa^{\underline{\alpha}} = \frac 1 2 (1-\tilde{\mathbf{\Gamma}})^{\underline{\alpha}}{}_{\underline{\beta}}  \kappa^{\underline{\beta}}=  \frac 1 4  \Big[(\tilde{v\!\!\!/}{} +\widehat{\tilde{s\!\!\!/}})
 \, ({v}\!\!\!/{}- \hat{s}\!\!\!/{}) \kappa \Big]^{\underline\alpha}\ee
where we have introduced the notation
\be\label{sG}
 {s\!\!\!/} = s^m{\Gamma}_m= s^mE_m^{\underline{a}}  {\Gamma}_{\underline{a}}\; , \qquad  \tilde{v}\!\!\!/ = v^m \tilde{\Gamma}_m = v^m E_m^{\underline{a}}  \tilde{\Gamma}_{\underline{a}} \; , \ee
 and used the normalized spacelike $s_m$ vector
 from \eqref{hsm:=},
 \bea\hat{s}^m = \frac {s^m} {\sqrt{-s^ls_l}}
 =g^{mn}\hat{s}_n
 \, , \qquad \hat{s}^m \hat{s}_m=-1\; . \qquad
\eea
The right hand side of eq. \eqref{kappa==} indicates that the
$\kappa$--symmetry projector is factorized into the product of two nilpotent matrices
\bea\label{vG-sG}
({v}\!\!\!/{}\pm  \hat{s}\!\!\!/{}) :=
(v\pm\hat{s})^{\underline{a}}\Gamma_{\underline{a}}=(v^m \pm \hat{s}^m)E_m^{\underline{a}}\Gamma_{\underline{a}} \; , \qquad \nonumber  \\ 
(\tilde {v}\!\!\!/{}\pm  \hat{\tilde s}\!\!\!/{})({v}\!\!\!/{}\pm  \hat{s}\!\!\!/{})=0\,. \qquad 
\eea
Their products in different orders provide us with orthogonal projectors
\be
{\cal P}_{+-}=\frac 1 4 (\tilde {v}\!\!\!/{}+  \hat{\tilde s}\!\!\!/{})({v}\!\!\!/{}- \hat{s}\!\!\!/{})\; , \qquad {\cal P}_{-+}=\frac 1 4 (\tilde {v}\!\!\!/{}-  \hat{\tilde s}\!\!\!/{})({v}\!\!\!/{}+ \hat{s}\!\!\!/{})
\ee
the first of which is exactly the $\kappa$--symmetry projector from  \eqref{kappa==},
\be {\cal P}_{+-} =\frac 1 2 (1-\tilde{\mathbf{\Gamma}}) \; . \qquad \ee

\subsection{Moving frame and kappa--symmetry} \label{mf-kappa}

Furthermore, it is important to note that the matrices \eqref{vG-sG} can be also factorized.
To find such a factorization, let us first notice that both vectors $v_m$  and $\hat{s}_m$ are normalized by construction
and hence are non-vanishing.  Then, the following eleven-- vector fields
\bea\label{hsa:=}
v^{\underline{a}}=v^mE_m^{\underline{a}}\qquad {\text and} \qquad \hat{s}^{\underline{a}}= \hat{s}^mE_m^{\underline{a}}\; , \qquad
\eea
obtained by contraction of the pull--back of the bosonic supervielbein with these vectors
($v^{\underline{a}}=i_vE^{\underline{a}}$,
$\hat{s}^{\underline{a}}= i_{\hat{s}}E^{\underline{a}}$)
are also normalized and orthogonal to each other
 \bea  \label{vhs=0} v^{\underline{a}}v_{\underline{a}}=1 \; , \qquad \hat{s}^{\underline{a}}\hat{s}_{\underline{a}}=-1 \, , \qquad \hat{s}^{\underline{a}} v_{\underline{a}}=0\; .  \qquad
\eea
\begin{widetext}
This set can be completed to form a $D=11$ Lorentz group valued matrix 
\be\label{vsu=in}
 \left( v_{\underline{a}} , u_{\underline{a}}^I,  \hat{s}_{\underline{a}} \right)= \left(\frac 1 2 \left( u_{\underline{a}}^{\#}+u_{\underline{a}}^{=}  \right) , u_{\underline{a}}^I, \frac 1 2\left(  u_{\underline{a}}^{\#}-u_{\underline{a}}^{=}\right) \right) \quad \in\quad SO(1,10)\; .
\ee
Besides the 11 vectors  $v$ and $\hat{s}$ of \eqref{hsa:=}, or their light-like linear combinations
\be\label{u--=v-hs}
u_{\underline{a}}^{=}:= (v-\hat{s})_{\underline{a}}\; , \qquad  u_{\underline{a}}^{\#} \equiv (v+\hat{s})_{\underline{a}} \; ,
\ee
this {\it moving frame matrix} contains nine  spacelike unit vectors $u_a^I$ orthogonal to both of that and between themselves.
\end{widetext}

Resuming, the above defined moving frame vectors
\footnote{They were also called light-cone harmonic variables \cite{Sokatchev:1985tc,Sokatchev:1987nk} and vector Lorentz harmonic variables (vector harmonics) \cite{Bandos:1990ji,Bandos:1990mw}. Notice that in the case of tensionless extended objects, the relevant variables are that used in superparticle and superstring cases \cite{Delduc:1991ir,Galperin:1991gk,Bandos:1992np,Galperin:1992pz,Bandos:1996ju,Bandos:2007wm} rather then those used to describe tensionfull super--p--branes with p$>$ 2 in its spinor moving frame formulation \cite{Bandos:1992hu,Bandos:1994eu} and in the frame of superembedding approach \cite{Bandos:1995zw,Sorokin:1999jx,Bandos:2023web}.
}
obey the set of relations
\bea\label{u--u--=0}
&& u_{\underline{a}}^{=} u^{\underline{a} =}=0\; , \qquad u_{\underline{a}}^{\#}u^{{\underline{a}}\#}=0 \; , \qquad  u_{\underline{a}}^{=}u^{{\underline{a}}\#} =2\; , \qquad
\\ \label{u--uI=0} && u_{\underline{a}}^{=} u^{{\underline{a}}I}=0\; , \qquad u_{\underline{a}}^{\#}u^{{\underline{a}}I}=0 \; , \qquad  u_{\underline{a}}^{I}u^{{\underline{a}}J} =-\delta^{IJ}\;  \qquad
\eea
which, if completed by
\be\label{epu1-u9=} \epsilon^{\underline{a}\underline{b}_1\ldots \underline{b}_9\underline{c}}v_{\underline{a}}u_{\underline{b}_1}{}^{J_1}\ldots u_{\underline{b}_9}{}^{J_9}\hat{s}_{\underline{c}} = \epsilon^ {J_1\ldots J_9}\, \ee
is equivalent to
\eqref{vsu=in}.

This vector moving frame in its turn  can be used to define a spinor moving frame following \cite{Bandos:1992np,Bandos:2007wm} and references therein. We are going to briefly describe this.

\subsection{Spinor moving frame} \label{sec=smf}

The spinor moving frame can be described by
$Spin(1,10)$ valued matrix
\be\label{VinSpin11}
V_{\underline{\alpha}}^{(\underline{\beta})} = \left(\begin{matrix} v_{\underline{\alpha}}^{ +q}\; ,  \; v_{\underline{\alpha}}^{\; -q} \end{matrix}\right)\; \in\; Spin(1,10)\;  \qquad
\ee
composed of two 16-dimensional sets of  $11D$ bosonic spinors. In addition to 11D Majorana spinor index ${\underline{\alpha}}=1,...,32$,
the spinor moving frame variables  $v_{\underline{\alpha}}^{+q}$ and $v_{\underline{\alpha}}^{-q}$ (also called spinor Lorentz harmonics or just Lorentz harmonics) carry SO(9) spinor index (i.e. Spin(9) index) $q=1,...,16$.

These {\it spinor moving frame variables} can be considered as a kind of square roots of the light--like vectors \eqref{u--=v-hs} as they obey the following Cartan-Penrose-type relations
\bea\label{v-s=v-v-}
 (\hat{v}\!\!\!/{}- \hat{s}\!\!\!/{})_{\underline{\alpha}\underline{\beta}}= 2 v_{\underline{\alpha}}^{-q}v_{\underline{\beta}}^{- q}\; , \quad (\tilde{v\!\!\!/}{} - \widehat{\tilde{s\!\!\!/}}) ^{\underline{\alpha}\underline{\beta}} = 2 v_{ {q}}^{-\underline{\alpha}}v_{ {q}}{}^{\!\!-\underline{\beta}}
 \, , \qquad
 \\ \nonumber \\  \label{v-s=v-Gv-}
 v^{-  {q}}\tilde{\Gamma}{}^{\underline{a}}v^{- {p}}=  (v-\hat{s})^{\underline{a}}\delta_{ qp}= v^-_{{q}}\Gamma^{\underline{a}}v^-_{ {p}}\; \qquad
\eea
and
\bea\label{v+s=v+v+}
 (\hat{v}\!\!\!/{}+ \hat{s}\!\!\!/{})_{\underline{\alpha}\underline{\beta}}= 2 v_{\underline{\alpha}}^{+q}v_{\underline{\beta}}^{+q}\; , \quad (\tilde{v\!\!\!/}{} + \widehat{\tilde{s\!\!\!/}}) ^{\underline{\alpha}\underline{\beta}} = 2 v_{ {q}}^{+\underline{\alpha}}v_{ {q}}{}^{+\underline{\beta}}
 \, , \qquad
 \\ \nonumber \\  \label{v+s=v+Gv+}
 v^{+ {q}}\tilde{\Gamma}{}^{\underline{a}}v^{+ {p}}=  (v-\hat{s})^{\underline{a}}\delta_{ qp}= v^+_{{q}}\Gamma^{\underline{a}}v^+_{ {p}}\; \qquad 
\eea
The bosonic spinor variables with upper
Majorana spinor indices, $v_{ {q}}^{\mp\underline{\alpha}}$, which are also used above,  are defined  by
\begin{eqnarray}
\label{V-1=CV-A} v_{q}^{+ {\underline{\alpha}}}  =  i C^{{\underline{\alpha}}{\underline{\beta}}}v_{{\underline{\beta}} q}^{\; +  }   \, ,
\qquad  v_{q}^{- {\underline{\alpha}}}  =  -i C^{{\underline{\alpha}}{\underline{\beta}}}v_{{\underline{\beta}} q}^{\; -  }   \, . \qquad
 \end{eqnarray}
This non-standard definition  makes them the blocks of the matrix inverse to the spinor moving frame matrix \eqref{VinSpin11}; this is to say, they obey
\be\label{v+v+=0}
v_{ {q}}^{+\underline{\alpha}}v_{\underline{\alpha}}^{+{p}}= 0\; , \qquad v_{ {q}}^{+\underline{\alpha}}v_{\underline{\alpha}}^{- {p}}=\delta_{{q}{p}}\; , \qquad v_{ {q}}^{-\underline{\alpha}}v_{\underline{\alpha}}^{- {p}}=0\; ,  \qquad \ee
and, hence,
\bea\label{I=vv+vv}
\delta_{\underline{\beta}}{}^{\underline{\alpha}}= v_{\underline{\beta}}^{+ {q}}v_{ {q}}^{-\underline{\alpha}} + v_{\underline{\beta}}^{-{q}}v_{{q}}^{+\underline{\alpha}} \; .  \qquad
\eea

The orthogonal moving frame vectors
$u_{\underline{a}}^I$ are composed from  both types of harmonics   $v_{ \underline{q}}^{+\underline{\alpha}}$, $v_{ \underline{q}}^{-\underline{\alpha}}$ as these obey, besides \eqref{v-s=v-v-}--\eqref{v+s=v+Gv+}, the following constraints
\bea \label{uIs=v+v-=11D}
 v_{{q}}^- \tilde {\Gamma}_{{\underline{a}}} v_{{p}}^+=\; u_{ {\underline{a}}}^{I} \gamma^I_{q{p}}, \quad
 u_{{\underline{a}}}^{I} {\Gamma}^{{\underline{a}}}_{{\underline{\alpha}}{\underline{\beta}}} =2 v_{( {{\underline{\alpha}}}|{q} }{}^- \gamma^I_{qp}v_{|{{\underline{\beta}}}){p}}{}^{+} , \qquad
 \\
 \label{uIts=v+v-=11D}
 v_{{q}}^- {\Gamma}_{{\underline{a}}} v_{{p}}^+=-u_{ {\underline{a}}}^{I} \gamma^I_{q{p}} , \quad
 u_{{\underline{a}}}^{I} \tilde{\Gamma}^{{\underline{a}}\underline{\alpha}\underline{\beta}} =-2 v_q^{-(\underline{\alpha}} \gamma^I_{qp}v_p^{+\underline{\beta})} . \qquad
 \eea
Here $\gamma^I_{qp}$ are SO(9) Dirac matrices obeying
\be \label{gIgJ+=I}\gamma^{I}_{qr}\gamma^{J}_{rp} +\gamma^{J}_{qr}\gamma^{I}_{rp} =2\delta^{IJ}\delta_{qp}\; , \qquad \gamma^I_{qp}=\gamma^I_{p{q}}, \qquad \gamma^I_{qq}=0\; . \qquad  \ee

The constraints \eqref{v-s=v-Gv-}--\eqref{v+s=v+Gv+} and  \eqref{uIs=v+v-=11D} can be obtained as the conditions of the Lorentz invariance of 11D gamma matrices specified for the case when Spin(1,10) valued spinor moving frame matrix \eqref{VinSpin11} and stating that it is the double covering of the SO(1,10) valued  moving frame matrix \eqref{vsu=in}. More details on spinor moving frame variables can be found   in Appendix \ref{app=movingframe} as well as  in e.g. \cite{Bandos:2007wm,Bandos:2022dpx} and references therein.

As an example of application of the spinor moving frame formalism to the case of tensionless M5-brane described by the action \eqref{SM5-T0=}, let us consider the equations for the  coordinate functions, Eqs. \eqref{fEq=} and \eqref{bEq=}.
Using the definition of the moving frame vectors in Eqs. \eqref{hsa:=}, \eqref{u--=v-hs} and  the constraint
\eqref{v-s=v-v-} we can write the fermionic equation \eqref{fEq=} in the following  irreducible 16 component form
\begin{eqnarray}
\label{fEq=-q} (v^m -\, \hat s^m) \partial_m \Theta^{\underline{\alpha}}\,
v_{\underline{\alpha}q}^{\;\; -}=0\;   \qquad
\end{eqnarray}
and the bosonic equation \eqref{bEq=} as
\begin{eqnarray}
\label{bEq==u--}
\partial_m (\sqrt{-g}\,\sqrt{-s^ls_l}\,(v^m -\, \hat s^m)u^=_{\underline{a}})=0
\; .  \qquad \end{eqnarray}
As we will see below, this equation can be split into the nine and one independent Lorentz invariant equations. But first, let us discuss the gauge fixing of the local fermionic $\kappa$--symmetry which can be performed using the spinor moving frame formalism.

\subsection{kappa--symmetry gauge fixing} \label{smf-kappa}

Using \eqref{v-s=v-v-} and \eqref{v+s=v+v+} we can resolve the constraint \eqref{kap=-bGkap}  on the $\kappa$--symmetry parameters  as follows
\be\label{kappa=} \kappa^{\underline{\alpha}} = v_{ {q}}^{-\underline{\alpha}} \kappa^{+ {q}}\; , \qquad
 \ee
which implies that we have expressed the constrained
32-component Majorana-Weyl spinor $\kappa^{\underline{\alpha}} $ in terms of the independent 16--component spinor  of the $SO(9)$ group $\kappa^{+ {q}}= \kappa^{\underline{\alpha}} v_{\underline{\alpha}}{}^+_{ {q}}$. Then eq. \eqref{vTh=ep+kap} takes the form

\be\label{vTh=ep+kqvq}
\delta\Theta^{\underline{\alpha}} = \epsilon^{\underline{\alpha}} + v_{{q}}^{-\underline{\alpha}} \kappa^{+ {q}} \; .
\ee

The natural gauge fixing condition for the fermionic gauge transformations with parameter $ \kappa^{+ {q}}(\xi)$ is then
\be\label{Th+q=0}
\Theta^{+\underline{q}(\mathfrak{v})}:= \Theta^{\underline{\alpha}}  {\mathfrak v}_{\underline{\alpha}\underline{q}}^{\;\; +}=0 \;
\ee
where ${\mathfrak v}_{\underline{\alpha}\underline{q}}^{\;\; +}$ is an element of a constant spinor frame,
\be
d{\mathfrak v}_{\underline{\alpha}\underline{q}}^{\;\; +}=0\; ,
\ee
which is 'sufficiently close' to our spinor moving frame constructed on $v_{\underline{a}}$ and  $s_{\underline{a}}$ vectors in the following sense.
We should require that the matrix $v_{\underline{q}}^{-\underline{\alpha}} {\mathfrak v}_{\underline{\alpha}\underline{q}}^{\;\; +} $ is invertible, i.e. that
\be\label{det=v-fv}
{\rm det} (v_{\underline{p}}^{-\underline{\alpha}} {\mathfrak v}_{\underline{\alpha}\underline{q}}^{\;\; +} )\not=0\; .
\ee

Clearly, this implies that we are restricting ourselves to field configurations
close enough  to (in the neighborhood of) some fixed 'constant background' one, which is described by a constant timelike unit vector
$\hat{s}{}^{(0)\underline{a}}= \frac {s^{(0)m} E_m^{(0)\underline{a}}}{\sqrt{-s_n^{(0)}g^{(0)mn}s^{(0)}_n}} $ and a fixed  timelike $v^{(0)\underline{a}}= v^{(0)m} E_m^{(0)\underline{a}} $ giving rise to constant vector frame
\be\label{fu--fu++fuI}
{\mathfrak u}^=_{\underline{a}}=\hat{s}{}^{(0)}_{\underline{a}} -v^{(0)}_{\underline{a}} , \qquad {\mathfrak u}^\#_{\underline{a}}=\hat{s}{}^{(0)}_{\underline{a}} +v^{(0)}_{\underline{a}}  , \qquad {\mathfrak u}^J_{\underline{a}}\;
\ee
obeying the counterpart of the constraints  \eqref{u--u--=0}, \eqref{u--uI=0}.

The square roots of the corresponding  constant vector frame matrix
\eqref{fu--fu++fuI}
is then given by a constant spinor frame matrix consisting of constant
32$\times$16 blocks $\mathfrak{v}_{\underline{\alpha}}^{\pm q}$, obeying a counterpart of the constraints \eqref{VinSpin11}, and hence of
\eqref{v-s=v-v-}--\eqref{v+s=v+Gv+} and \eqref{uIs=v+v-=11D}
involving also the constant frame vectors \eqref{fu--fu++fuI}.

In the generic case any moving frame can be related with other, or with fixed frame, by local Lorentz transformations the general form of which can be found in \cite{Bandos:2017eof} and in Appendix \ref{app:v-fv} of the present paper (particularly in  Eqs. \eqref{u--=KuI=11}--\eqref{v-1+=fv-1+}). At the present stage we need only two of such relations
\begin{eqnarray}\label{v-1-=fv-1-}
v_{q}^{-\underline{\alpha}} &=& e^{-\beta} {\cal O}_{{q}{p}} \left(
\mathfrak{v}_{{p} }^{ -\underline{\alpha}}- \frac {1}{2} K^{=I} \gamma^I_{p{p}'} \mathfrak{v}_{{p}' }^{+\underline{\alpha}}
\right)
  \; ,\nonumber  \\ {} \\ 
\label{u--=fu--+}
 u^=_{\underline{a}} &:=& (v_{\underline{a}}-\hat{s}_{\underline{a}})= e^{-2\beta}\left(\mathfrak{u}^=_{\underline{a}} + \frac 1 4 (\vec{K}^{=} )^2 \mathfrak{u}^\#_{\underline{a}}  + K^{=I}\mathfrak{u}^I_{\underline{a}} \right)
 \; .\nonumber  \\ {}
\end{eqnarray}
Here  which $\beta$ parametrizes $SO(1,1)$ subgroup of $SO(1,10)$,
 ${K}{}^{=J}$ parametrizes the coset of the spacetime Lorentz group
 isomorphic to celestial sphere ${\bb S}^{9}\;$, \footnote{
This is $SO(1,10)/H$, where $H$ is the parabolic subgroup of the 11D Lorentz group, see \cite{Delduc:1991ir,Galperin:1991gk} for the discussion.} and
${\cal O}_{pq}={\cal O}^{-1}_{qp}$ is the Spin(9) valued matrix (notice that Spin(9)$\subset$ SO(16)). All these parameters become functions depending on $\xi^m$ when we consider a relation of the (spinor) moving frame fields and the constant reference (spinor)  frame,
\bea
\beta=\beta (\xi)\; , \qquad {K}{}^{=J}={K}{}^{=J}(\xi)\; , \qquad \nonumber \\ {\cal O}_{pq}={\cal O}_{pq}(\xi) \; \in\; Spin(9) \; \subset SO(16)\; .
\eea

Using Eqs.  \eqref{v-1-=fv-1-} and  \eqref{u--=fu--+}, we find that the residual $\kappa$--symmetry transformations preserving the gauge \eqref{Th+q=0} has the parameter expressed through one of the projections of the constant spacetime supersymmetry parameter on constant spinor frame,
\be
\epsilon^{\mp q (\mathfrak{v})}:= \epsilon^{\underline{\alpha}} \mathfrak{v}^{\;\; \mp}_{\underline{\alpha}q}\; , \qquad
\ee
by
\be \label{kap=ep}
\kappa^{+q}= - e^{\beta(\xi) }
{\cal O}_{qp}(\xi)\epsilon^{+p (\mathfrak{v})}
\; .  \qquad \ee
The nonvanishing projection of the fermionic coordinate functions is transformed by both projections of the supersymmetry transformations,
 \be\label{varTh-q=}
\delta \Theta^{-q(\mathfrak{v})}  = \delta \Theta^{\underline{\alpha}} \mathfrak{v} _{\underline{\alpha}q}^{\; \; - } = \,\epsilon^{-q (\mathfrak{v})}+\, \frac 1 2\, K^{=I} \gamma^I_{qp} \epsilon^{+p (\mathfrak{v})} \; .
\ee
Notice the absence of any contribution of $Spin(9)$-valued matrix field ${\cal O}_{pq}$ in this transformation.

Actually,  $Spin(9)$ transformations  is a gauge symmetry of our construction. Indeed, these  are related to SO(9) rotation of the spacelike unit moving frame vectors $u_{\underline{a}}^I$ by
\begin{eqnarray}\label{OIJ=OGO}
 {\cal O}_{qr} {\cal O}_{{p}{s}} \gamma^{I}_{rs}= \gamma^{J}_{qp} {\cal O}^{JI}\; .
   \qquad
\end{eqnarray}
This implies that we are allowed to choose the gauge in which ${\cal O}_{{p}{q}}$ and  ${\cal O}^{IJ}$ are given by 16$\times 16$ and $9\times 9$ unity matrices. We however, pefer to use this symmetry in a different way (see below).

Now, observing that \be\label{e-2b=} e^{-2\beta}=
\frac 1 2 \mathfrak{u}^{\#}_{\underline{a}} (v-\hat{s})^{\underline{a}}\; , \qquad \ee
we can write Eq. \eqref{kap=ep} in the form of
\be \label{kap==ep}
\kappa^{+p}= - \, e^{-\beta (\xi)}
{\mathcal O}_{pq}\epsilon^{+q \mathfrak{v}}\,= - \,
{\mathcal O}_{pq}\epsilon^{+q \mathfrak{v}}\, \frac 1 {\sqrt{(v-\hat{s})^{\underline{a}}\, \mathfrak{u}^{\#}_{\underline{a}} /2\,}}\,
\; .  \qquad \ee
Similarly, we find
\bea\label{K--I=}
K^{=I} &=& -e^{2\beta}\, (v-\hat{s})^{\underline{a}}\, \mathfrak{u}_{\underline{a}}^I=  - 2 \frac { (v-\hat{s})^{\underline{a}}\, \mathfrak{u}_{\underline{a}}^I} {(v-\hat{s})^{\underline{b}}\, \mathfrak{u}^{\#}_{\underline{b}}}\;
\eea
and more complex expressions of these type  for $K^{\# I}(\xi) $ and ${\mathcal O}^{IJ}(\xi)$ the explicit form of which is not necessary for the discussion below.

Eqs. \eqref{e-2b=} and \eqref{K--I=} make explicit that the supersymmetry  transformations of the Goldstone fermion of tensionless M5--brane \eqref{varTh-q=} become field dependent in the gauge \eqref{Th+q=0}. In particular the r.h.s. of Eq.  \eqref{varTh-q=} depend on the worldvolume gauge field strength used to construct  the spacelike vector $\hat{s}$, (see Eqs. \eqref{sm:=}, \eqref{hsm:=} and \eqref{hsa:=}).

Below we will mainly use
the more economic description in terms of composite fields
$K^{=I}(\xi)$ and $\beta (\xi)$, and just keep in mind their explicit form of Eqs.  \eqref{K--I=} and \eqref{e-2b=}.

\medskip

The  equations of motion for the fermionic coordinate function, Eq.\eqref{fEq=-q}, in the gauge \eqref{Th+q=0} acquire the form
\be
\label{fEq=Th-q=}
(v^m -\, \hat s^m) \partial_m \Theta^{-q\mathfrak{v}}\,
=0\; .  \qquad
\ee

As far as the equation \eqref{bEq==u--} for the bosonic coordinate functions is concerned, contracting it with
constant moving frame vectors \eqref{fu--fu++fuI} we will find that it is split into  one plus nine independent equations
\bea\label{bEq=--}
\partial_m (\sqrt{-g}\,\sqrt{-s^ls_l}\,(v^m -\, \hat s^m)e^{-2\beta} )=0\; , \qquad \\ \label{bEq=dKI}
(v^m -\, \hat{s}^m)\partial_m  K^{=I} =0\; . \qquad
\eea
Below we will come back to these equations after gauge fixing the diffeomorphism symmetry.

\subsection{Supersymmetry after partial gauge fixing }

The supersymmetry transformations  of bosonic coordinate functions   in the gauge \eqref{Th+q=0} reads
\bea
\delta X^{\underline{a}} &=&  i \Theta^{-q(\mathfrak{v})}\left(\,\epsilon^{+p \mathfrak{v}} +\frac {1} {2} K^{=I} \gamma^I_{qp}   \,\epsilon^{+p \mathfrak{v}} \right)\,{\mathfrak{u}}^{\#\underline{a}}\nonumber \\ && \hspace{2cm} - 2i \Theta^{-q(\mathfrak{v})} \gamma^I _{qp} \epsilon^{+p \mathfrak{v}} \, {\mathfrak{u}}^{I \underline{a}} \; . \qquad
\eea
They  can be written in terms of  projections to the constant frame \eqref{fu--fu++fuI},
\be
X^{=}= X^{\underline{a}}  \mathfrak{u}_{\underline{a}} ^{=}\; , \qquad X^{\#}= X^{\underline{a}}  \mathfrak{u}_{\underline{a}} ^{\#}\; , \qquad X^{I}= X^{\underline{a}}  \mathfrak{u}_{\underline{a}} ^{I}\;  \qquad
\ee
as follows:
\bea\label{susy=X++}
\delta X^{\#} &=& 0\; , \qquad \\ \label{susy=X--}
\delta X^=  &=& 2i \Theta^{-q(\mathfrak{v})}\left(\,\epsilon^{-q \mathfrak{v}} -\frac {1} {2} K^{=I} \gamma^I_{qp}   \,\epsilon^{+p \mathfrak{v}} \right)\; , \qquad \\  \label{susy=XI} \delta X^I  &=& 2i \Theta^{-q(\mathfrak{v})} \gamma^I _{qp} \epsilon^{+p \mathfrak{v}} \; .  \qquad
\eea
Notice that the r.h.s. of  \eqref{susy=XI} has appeared as a sum of two terms, one of which comes from original contribution of spacetime supersymmetry transformations and the second-- from the $\kappa$--symmetry. In the gauge \eqref{Th+q=0} this latter gives the  contribution equal to the first term because  the parameter of $\kappa$--symmetry is expressed in terms of supersymmetry parameter by \eqref{kap=ep}.

\bigskip

To conclude this section, let us discuss the gauge fixed form of the pull--back of the bosonic supervielbein, $E^{\underline{a}}$, and its properties under supersymmetry.

In the gauge \eqref{Th+q=0}, the projections of the pull--back of the bosonic supervielbein form on the reference constant frame simplify to
\bea\label{E++=dX++}
E^{\# \, ({\mathfrak{u}})} &=& dX^{\# \, ({\mathfrak{u}})}\; , \qquad
\\ \label{E--=dX--+}
E^{= \, ({\mathfrak{u}})}&=& dX^{= \, ({\mathfrak{u}})} - 2id\Theta^{-\underline{q}(\mathfrak{v})}\, \Theta^{-\underline{q}(\mathfrak{v})}\; , \qquad \\
 \label{EI=dXI}
E^{I \, ({\mathfrak{u}})}&=& dX^{I \, ({\mathfrak{u}})}\; .  \qquad
\eea
Notice that after gauge fixing \eqref{Th+q=0} these one-forms are not  invariant under spacetime supersymmetry transformations as these induce residual $\kappa$--symmetry  with parameter \eqref{kap=ep} and, under $\kappa$--symmetry, the bosonic supervielbein is not invariant  but transforms through fermionic supervielbein.
The straightforward calculations show that, after gauge fixing,
\be
\delta_{\epsilon} E^{ \underline{a}}=-2i d\Theta^{-q(\mathfrak{v})} \gamma^I _{qp} \epsilon^{+p \mathfrak{v}} \mathfrak{u}^{I \underline{a}} -i d\Theta^{-q(\mathfrak{v})}K^{=I} \gamma^I _{qp} \epsilon^{+p \mathfrak{v}} \mathfrak{u}^{\# \underline{a}}
\ee
and hence
\bea
\delta_{\epsilon} E^{\# \, ({\mathfrak{u}})}&=& 0 \; , \qquad
\\
\delta_{\epsilon}  E^{= \, ({\mathfrak{u}})}&=& -2i d\Theta^{-q(\mathfrak{v})}K^{=I} \gamma^I _{qp} \epsilon^{+p \mathfrak{v}}\; , \qquad
\\
\delta_{\epsilon}  E^{I \, ({\mathfrak{u}})} &=& 2i d\Theta^{-q(\mathfrak{v})} \gamma^I _{qp} \epsilon^{+p \mathfrak{v}} \; .  \qquad
\eea

\subsection{Gauge fixing of the worldvolume diffeomorphism symmetry }

Using the definition of moving frame \eqref{vsu=in} build on $(v\pm s)^{\underline{a}}=(v\pm s)^n E_n^{\underline{a}}$ vector fields and relation between this and the constant frame, in  particular Eq.  \eqref{u--=fu--+},
we can find that in the gauge \eqref{Th+q=0} (in which the bosonic supervielbein has the form of \eqref{E++=dX++}--\eqref{EI=dXI})
\bea\label{v-hs=gaugef}
(v-\hat{s})^m \partial_m X^{\#}&=& 2e^{-2\beta}\; , \qquad \nonumber \\  (v-\hat{s})^m \partial_m X^{I}&=&-  e^{-2\beta}K^{=I}\; , \qquad  {}  \\ \label{v-hs=gaugef--}
(v-\hat{s})^m \partial_m X^{=} &-& 2i(v-\hat{s})^m \partial_m\Theta^{-\underline{q}}\, \Theta^{-\underline{q}}= \frac 1 2 e^{-2\beta}\,  (\vec{K}^{=})^2. \nonumber \\ {}
\eea
Here and below  we skip the $^{(\mathfrak{u})}$ and $^{(\mathfrak{v})}$ superscripts from the projections on the constant frame and  constant spinor frame as we will not use other projections.

Notice that the fermionic contribution to \eqref{v-hs=gaugef--} vanishes
on-shell due to the fermionic equations of motion \eqref{fEq=-q}. We have kept it
to maintain all the above set of equations, \eqref{v-hs=gaugef} and \eqref{v-hs=gaugef--}, to be valid off-shell.

\bigskip

\begin{widetext}
At this stage it is convenient, on one hand,  to split the set of 6 worldvolume coordinates as
\bea
\xi^m=(\xi^{++}, \xi^{--}, \xi^{\check{m}})\; , \qquad
\xi^{++}=\xi^{0}+\xi^{5}, \qquad   \xi^{--}=\xi^{0}-\xi^{5}\; , \qquad \xi^{\check{m}}=(\xi^{1}, \xi^{2}, \xi^{3}, \xi^{4})\qquad
\eea
and, on the other hand, to use the local SO(9) symmetry of our spinor moving frame construction to gauge away four of nine components of the 9-vector field $K^{=I}(\xi)= (K^{=\check{i}} (\xi), K^{=\check{I}} (\xi)) $:
\be\label{K--ch-i=0}
K^{=\check{i}} (\xi) =0 \qquad \Rightarrow \qquad K^{=I}(\xi) = \delta_{\check{J}}{}^I K^{=\check{J}} (\xi)) \; , \qquad
{\check{i}}=1,2,3,4 \; , \qquad {\check{I}}, {\check{J}}=5,6,7,8,9\; .
\ee

Making corresponding  SO(4)$\times$SO(5) splitting of the nanoplet of the bosonic coordinate functions,
\be
X^I (\xi) = (X^{\check{i}} (\xi), X^{\check{I}} (\xi)) \; , \qquad
{\check{i}}=1,2,3,4 \; , \qquad {\check{I}}=5,6,7,8,9\; ,
\ee
we find that the second relation in \eqref{v-hs=gaugef} splits in quadruplet of homogeneous equations and five relations expressing $K^{=\check{J}} (\xi)$ through derivatives of $X^{\check{I}} (\xi)$,
\bea\label{(v-s)dXi=0} (v-\hat{s})^m \partial_m X^{\check{i}}=0 \; , \qquad {\check{i}}=1,2,3,4 \; , \qquad \\ \label{(v-s)dXI=-KI} K^{=\check{I}}=-e^{2\beta} (v-\hat{s})^m \partial_m X^{\check{I}}\; , \qquad {\check{I}}=5,6,7,8,9\; . \eea

Now we are ready to fix the gauge with respect to all  six worldvolume diffeomorphism symmetries by
\bea\label{X++=xi++}
X^{\#}:= X^{ \underline{a}}(\xi) \mathfrak{u}^\#_{ \underline{a}} = \xi^{++}\; , \qquad \\ \label{X--=xi--}
X^{=}:= X^{ \underline{a}}(\xi) \mathfrak{u}^=_{ \underline{a}} = \xi^{--}\; , \qquad
\\ \label{Xi=xii}
X^{\check{i}}:= X^{ \underline{a}}(\xi) \mathfrak{u}^I_{\underline{a}} \delta^{\check{i}}_I=  \xi^{\check{m}} \delta_{\check{m}}{}^{\check{i}} . \quad
\eea

In this gauge Eqs.  \eqref{v-hs=gaugef} and \eqref{v-hs=gaugef--} acqure the form
\bea\label{v-hs=gauge=}
(v-\hat{s})^{++}= 2e^{-2\beta}\; , \qquad (v-\hat{s})^{\check{i}} =0\; , \qquad \\  \label{v-hs=gauge=--} (v-\hat{s})^{--}=
\frac 1 2 (v-\hat{s})^{++} ({K}^{=\check{I}})^2+ 2i(v-\hat{s})^m \partial_m\Theta^{-\underline{q}}\, \Theta^{-\underline{q}}\; , \qquad
\\ \label{v-hs=gauge=cI}
(v-\hat{s})^m \partial_m X^{\check{I}}=-  e^{-2\beta}{K}^{=\check{I}} =- \frac 1 2 (v-\hat{s})^{++}K^{=\check{I}} \; . \qquad
\eea

\subsubsection{Gauge fixed equations of motion}

To simplify further the analysis, at least in the first stage of the study,  it is convenient to  use the fermionic equations of motion \eqref{fEq=-q} to reduce  \eqref{v-hs=gauge=--} to
\be\label{v-hs=gauge=--on} (v-\hat{s})^{--}=
\frac 1 4 (v-\hat{s})^{++} ({K}^{=\check{I}})^2\; .  \qquad\ee
A discussion on the off-shell gauge fixing can be found in Appendix \ref{app:off-shell}.

Clearly, in the purely bosonic case the first of equations \eqref{v-hs=gauge=} can be valid  provided $(v-\hat{s})^{++}>0$ and then Eq. \eqref{v-hs=gauge=--}
implies that also $(v-\hat{s})^{--}>0$. These relations  define  a kind of coordinate chart in which we have fixed the gauge. Of course in the presence of nilpotent contributions from fermionic fields, the above iniqualities does not make sense.

It is important that in the gauge \eqref{Th+q=0}, \eqref{X++=xi++}--\eqref{Xi=xii}
\bea
(v-\hat{s})^m \partial_m =
\frac 1 2 (v-\hat{s})^{++}\, \nabla_{++}\; , \qquad \\ \label{nabla=K}
 \nabla_{++}= \partial_{++}+ \frac 1 4 ({K}^{=\check{I}})^2\partial_{--} \, . \qquad
\eea
In particular, taking into account that $(v-\hat{s})^{++}\not= 0$, one finds that  \eqref{v-hs=gauge=cI} implies
\bea\label{K--I=d++XI+e=cI}
 \nabla_{++}  X^{\check{I}}:=
\left(\partial_{++}+ \frac 1 4 ({K}^{=\check{J}})^2\partial_{--}\right) X^{\check{I}} =-{K}^{=\check{I}}\; . \qquad
\eea
Instead of trying to solve this equation with respect to ${K}^{=\check{I}}$, we will use it below as a formal relation to analyze equations of motion. However,
such a solution certainly exists at least as a power series in the approximation of slowly varying fields, \be {K}^{=\check{I}}= - \partial_{++}X^{\check{I}} - \frac 1 4 \, (\vec{K}^{=\check{J}})^{2} \partial_{--}{X}^{=\check{I}}=- \partial_{++}X^{\check{I}} - \frac 1 4 \, \partial_{++}X^{\check{J}}\partial_{++}X^{\check{J}} \, \partial_{--} X^{\check{I}} + ... \; . \ee

The  fermionic and bosonic equations \eqref{fEq=Th-q=} and \eqref{bEq=dKI}  in the gauge \eqref{Xi=xii} read
\bea
\label{fEq=dTh-q=}
\nabla_{++} \Theta^{-q\mathfrak{v}}\, =\left(\partial_{++}+ \frac 1 4 ({K}^{=\check{J}})^2\partial_{--}\right) \Theta^{-q\mathfrak{v}}\, =0\; , \qquad \\
\label{bEq=dKI==}
\nabla_{++} K^{=\check{I}}\, =\left(\partial_{++}+ \frac 1 4 ({K}^{=\check{J}})^2\partial_{--}\right) K^{=\check{I}}\, =0\;   \qquad
\eea
and simplify when we use (non-holonomic) derivative
\bea\label{nabla++==} \nabla_{++} & =&
\partial_{++}+ \frac 1 4 ({K}^{=\check{J}})^2\partial_{--} =
\partial_{++}+ \frac 1 4 (\nabla_{++}{X}^{\check{J}})^2\partial_{--} = \nonumber \\ &=& \partial_{++}+ \frac 1 4 \left( (\partial_{++}{X}^{=\check{J}})^2+\ldots \right) \partial_{--} \; .
\eea

Using this and \eqref{K--I=d++XI+e=cI} one  immediately finds that \eqref{bEq=dKI==} is equivalent to
\be\label{d2XI=0}
\nabla_{++}\nabla_{++}  X^{\check{I}}=0\; .
\ee
Notice that, due to \eqref{bEq=dKI==},
\bea
\nabla_{++}\nabla_{++}= \partial_{++}\partial_{++}+ \frac 1 2 ({K}^{=\check{J}})^2\partial_{++}\partial_{--}+ \frac 1 {16}
(({K}^{=\check{J}})^2)^2\partial_{--}\partial_{--}\; .
\eea
Eqs. \eqref{d2XI=0} and \eqref{fEq=dTh-q=} resemble superstring or null-superstring equations in light-cone gauge or in spinor moving frame approach, in which $({K}^{=\check{J}})^2$ plays the role of Beltrami differentials. However, in distinction to  that, Eqs. \eqref{d2XI=0} and \eqref{fEq=dTh-q=} are highly nonlinear due to the field dependent contributions to the derivative $\nabla_{++}$   \eqref{nabla++==}.

The gauge fixed from of Eq. \eqref{bEq=--} is a bit more complicated:
\be\label{bEq=--=}
0 =\nabla_{++} (\sqrt{-g}\,\sqrt{-s^ls_l}\,((v -\, \hat{s})^{++})^2)+ \frac 1 2 \sqrt{-g}\,\sqrt{-s^ls_l}\,((v -\, \hat{s})^{++} )^2\, {K}^{=\check{J}}\partial_{--}{K}^{=\check{J}}\; . \qquad
\ee
As far as nonlinear self-duality equations \eqref{ivH3=-cV2}  are concerned,
\begin{eqnarray}\label{ivH3=-cV2==}
 H_{mnk}v^k=-\frac {\sqrt{-g}}{2\sqrt{-s^ss_s} } \epsilon_{mnrlpq} s^{r}v^l \tilde{H}^{pq} =-\frac {\sqrt{-g}}{2} \epsilon_{mnrlpq} \hat{s}^{r}v^l \tilde{H}^{pq}\; ,
\end{eqnarray}
 their gauge fixed version does not look especially simple.  Notice only that it implies

\begin{eqnarray}\label{sivH3==0}
 H_{mnk}v^ks^{n}=0 \;.
\end{eqnarray}
 Neither simplification is seen in the  Lagrangian equations for the 2--form gauge field which follow directly from the PST action
\be\label{b2=Eqs}
\partial_m (\sqrt{-g} (v-\hat{s})^{[m} \tilde{H}{}^{nl]} )=0\;
\ee
 (see \cite{Bandos:2026pdg} for the discussion of its properties and the derivation of \eqref{sivH3==0} from it).

\bigskip

\subsection{Residual symmetries of the gauge fixed description of tensionless M5--brane}

The expression \eqref{kap=ep} describe the residual $\kappa$--symmetry transformations induced by spacetime supersymmetry transformations. However, this can be induced also by the Lorentz group. Indeed, using the complete form including also Poincare and scaling symmetry transformations of fermionic coordinate functions before the gauge fixing, Eq. \eqref{vTh=sPoin+scal}, one finds\footnote{This calculation uses  the constant spinor moving frame counterparts of the constraints \eqref{u==v-v-=11D}--\eqref{uIs=v+v-=11D=}.}
\be \label{kap=ep++L}
\kappa^{+q}= -  e^{\beta}\mathcal{O}_{qp}(\xi) \left(\epsilon^{+p(\mathfrak{v})}+\frac 1 2 L^{\# I} \Theta^{-r(\mathfrak{v})} \gamma^I_{rp} \right)
\; .  \qquad \ee
Thus, the transformations of e.g. Goldstone fermion field in the gauge reads
 \bea\label{varTh-q==}
\delta \Theta^{-q(\mathfrak{v})}  = \delta \Theta^{\underline{\alpha}} \mathfrak{v} _{\underline{\alpha}q}^{\; \; - } &=& \,\epsilon^{-q (\mathfrak{v})}+\, \frac 1 2\, K^{=I} \left( \epsilon^{+p (\mathfrak{v})} \gamma^I_{pq}+ \frac 1 2\, L^{\# J}\Theta^{-p(\mathfrak{v})}(\gamma^I\gamma^J)_{pq}\right)-\qquad \nonumber \\
&&- \frac 1 4\,  \left( L^{\# =} \Theta^{-q(\mathfrak{v})}
- L^{IJ} \Theta^{-p(\mathfrak{v})}
(\gamma^{IJ})_{pq}\right)\; . \qquad
\eea
This reduces to \eqref{varTh-q=} when the parameter of Lorentz transformations $L^{\underline{a}\underline{b}}$, and hence its projections $L^{\# =}$, $L^{\# I}$ and $L^{IJ}$ are set to zero.

In the gauge \eqref{Th+q=0}  the transformations of the bosonic coordinate functions under spacetime symmetries, including supersymmetry,     acquire the form (see Appendix \ref{derivation=vXa} for derivation of this)
\bea
\label{vX=gaugeTh}
\delta X^{\underline{a}}
 &=& a^{\underline{a}} +  X^{\underline{b}}
L_{\underline{b}}{}^{\underline{a}}+\lambda X^{\underline{a}} +i\Theta^{-q(\mathfrak{v})}
\left(
\mathfrak{u}^{\# \underline{a}}\epsilon^{-q(\mathfrak{v})}  + \mathfrak{u}^{I \underline{a}}\gamma^I_{qp}\epsilon^{+p(\mathfrak{v})}
\right)
+ \qquad \nonumber \\ && \qquad + i\left(\Theta^{-q(\mathfrak{v})}\gamma^I_{qp}\epsilon^{+p(\mathfrak{v})}+ \frac 1 2L^{\# J} \Theta^{-q(\mathfrak{v})}\gamma^{IJ}_{qp} \Theta^{-p(\mathfrak{v})}  \right) \left(\mathfrak{u}^{I \underline{a}}- \frac 1 2 K^{=I}\mathfrak{u}^{\# \underline{a}}\right)
\; .  \qquad
\eea
This implies that the residual diffeomorphism transformations  which preserve the light-cone gauge of Eqs. \eqref{X++=xi++}--\eqref{Xi=xii} and \eqref{Th+q=0} act on the worldvolume coordinates $\xi^m=(\xi^{++}, \xi^{--}, \xi^{\check{i}})$ as follows
\bea  \label{res-v-xi++}
&& \delta \xi^{++}= a^{\#}+\, \xi^{++}\left(\lambda -\frac 1 2 L^{\#\, =}\right)+ \xi^{\check{i}}  L^{\#\, \check{i}}  + \; \qquad \nonumber \\ && {}\qquad + X^{\check{I}} (\xi) L^{\#\, \check{I}}\; , \qquad
\\ \label{res-v-xi--}
&& \delta \xi^{--}= a^{=} +\, \xi^{--} \left(\lambda +\frac 1 2 L^{\#\, =}\right) + \xi^{\check{i}}  L^{=\, \check{i}} +\; \qquad \nonumber \\ &&{}\qquad + X^{\check{I}}  L^{=\, \check{I}}+  2i \Theta^{-q(\mathfrak{v})} \epsilon^{-q \mathfrak{v}} +i \nabla_{++}X^{\check{I}} \Theta^{-q(\mathfrak{v})} \gamma^{\check{I}}_{qp}   \,\epsilon^{+p \mathfrak{v}}- \nonumber \\ &&{}\qquad + \frac i 2 \nabla_{++}X^{\check{I}}L^{\# J} \Theta^{-q(\mathfrak{v})}\gamma^{\check{I}J}_{qp} \Theta^{-p(\mathfrak{v})}   \, , \qquad
\\ \label{res-v-xii}
&& \delta \xi^{\check{i}}= a^{\check{i}}+\, \xi^{\check{j}}\left(\lambda \delta^{\check{j}\check{i}} - L^{\check{j}\check{i}}\right) + \frac 1 2 \xi^{++} L^{=\,\check{i}} + \frac 1 2 \xi^{--}L^{\# \check{i}} - \; \qquad \nonumber \\ && {}\qquad - X^{\check{J}} (\xi) L^{\check{J}\check{i}}- 2i \Theta^{-q(\mathfrak{v})}(\xi)\gamma^{\check{i}}_{qp}\epsilon^{+p \mathfrak{v}}- \frac i 2  L^{\# J} \Theta^{-q(\mathfrak{v})}\gamma^{\check{i}J}_{qp} \Theta^{-p(\mathfrak{v})}   \; , \qquad
\eea
where $\gamma^{\check{i}}=\delta^{\check{i}}_I \gamma^I$,
$\nabla_{++}$ is defined in \eqref{nabla++==} and we have used Eq. \eqref{K--I=d++XI+e=cI}

These transformations contain the contributions of the bosonic and fermionic coordinate fields which are collected in the second and the third lines of the above expressions.

The transformations of the bosonic and fermionic Goldstone fields of the tensionless M5--brane in the gauge \eqref{X++=xi++}--\eqref{Xi=xii} and \eqref{Th+q=0} read

\bea \label{res-v-XchI=}
{\delta} X^{\check{I}}&:=& X^{\check{I}\prime }(\xi^\prime) - X^{\check{I}}(\xi) = a^{\check{I}}+\, \xi^{\check{j}}\left(\lambda \delta^{\check{j}\check{I}} - L^{\check{j}\check{I}}\right) + \frac 1 2 \xi^{++} L^{=\,\check{I}} + \frac 1 2 \xi^{--}L^{\# \check{I}} - \; \qquad \nonumber \\ &&  - X^{\check{J}} (\xi) L^{\check{J}\check{I}}- 2i \Theta^{-q(\mathfrak{v})}(\xi)\gamma^{\check{I}}_{qp}\epsilon^{+p \mathfrak{v}}- \frac i 2  L^{\# J} \Theta^{-q(\mathfrak{v})}\gamma^{\check{I}J}_{qp} \Theta^{-p(\mathfrak{v})}\; \qquad
\eea
and
 \bea\label{varTh-q====}
{\delta} \Theta^{-q(\mathfrak{v})}   &:=&  \Theta^{-q(\mathfrak{v})\prime}(\xi^\prime)- \Theta^{-q(\mathfrak{v})\prime}(\xi) = \qquad \nonumber \\ &=& \,\epsilon^{-q (\mathfrak{v})}-\, \frac 1 2\, \nabla_{++}X^{\check{I}} \left( \epsilon^{+p (\mathfrak{v})} \gamma^{\check{I}}_{pq}+ \frac 1 2\, L^{\# J}\Theta^{-p(\mathfrak{v})}(\gamma^{\check{I}}\gamma^J)_{pq}\right)-\qquad \nonumber \\
&&- \frac 1 4\,  \left( L^{\# =} \Theta^{-q(\mathfrak{v})}
- L^{IJ} \Theta^{-p(\mathfrak{v})}
(\gamma^{IJ})_{pq}\right) \; , \qquad
\eea
where  $\gamma^{\check{I}}=\delta^{\check{I}}_J \gamma^J$.

\end{widetext}
If we are interested in the variation of field at the same worldvolume point, we have to use the expression
\bea \label{res-v-XchI}
\underline{\delta} X^{\check{I}}&:=& X^{\check{I}\prime }(\xi) - X^{\check{I}}(\xi) = {\delta} X^{\check{I}}
- \delta \xi^m \partial_m   X^{\check{I}}\;  \qquad
\eea
and
 \bea\label{varTh-q===}
\underline{\delta} \Theta^{-q}   &:=&  \Theta^{-q\prime}(\xi)- \Theta^{-q}(\xi) = {\delta} \Theta^{-q}   - \delta \xi^m \partial_m \Theta^{-q}\; . \qquad
\eea
However, if calculating the algebra of the (super)symmetries preserving the gauge with respect to $\kappa$--symmetry and diffeomorphisms, and taking the second variation to this end, one has to use  the variations \eqref{res-v-XchI=} and  \eqref{varTh-q====}\footnote{The reason is that the condition of preservation of, say, the gauge \eqref{Xi=xii} reads $\delta X^{\check{i}}(\xi):=
X^{\check{i}\prime}(\xi^\prime) - X^{\check{i}}(\xi)=\delta \xi^{\check{i}} =\xi^{\check{i}\prime} -\xi^{\check{i}}
$.}.

Notice that transformations of the bosonic coordinate field, Eq. \eqref{res-v-XchI=}, include the terms linear in the worldvolume coordinates,  which are collected in the first line together with constant shift $a^{\check{I}}$ indicating the Goldstone nature of
$X^{\check{I}}(\xi)$. This is not the case for fermionic Goldstone field transformations \eqref{varTh-q====} which do not include contributions proportional to the worldvolume coordinates.

In contrast, the variations at the same worldvolume point, \eqref{res-v-XchI} and \eqref{varTh-q===} in both cases contain contributions proportional to the coordinates coming from   $\delta \xi^m $  given in \eqref{res-v-xi++} -- \eqref{res-v-xii}.

The transformations of the bosonic and fermionic fields in \eqref{res-v-XchI=} and  \eqref{varTh-q===} clearly indicate that the spacetime supersymmetry with parameter $\epsilon^{-q}$ is realized nonlinearly with $\Theta^{-q}$ being its Goldstone field, while $\epsilon^{+q}$ supersymmetry is linearly realized and  acts on bosonic and fermionic field in the manner characteristic for a supermultiplet: bosonic field is transformed through fermionic one and the fermionic field transforms through the derivative of the bosonic field, though the corresponding derivative $\nabla_{++}$ is actually quite nonlinear in partial derivatives of the bosonic fields,
$\partial_{++}X^{\check{I}}$.

Notice that the Stueckelberg field describing the SO(9) transformations, ${\mathcal{O}}^{IJ}(\xi)$ which  relates the vectors $u^I_{\underline{a}}$ orthogonal to $(v\mp \hat{s})^m E_m^{\underline{a}}$ with the vectors of the constant frames, in particular with  $\mathfrak{u}^J_{\underline{a}}$, does not enter the residual diffeomorphism symmetry transformations \eqref{res-v-xi++}--\eqref{res-v-xi++}. These is natural as SO(9) is a manifest gauge symmetry of our construction of moving frame on the basis of two light-like vectors $(v\pm \hat{s})^{\underline{a}}= (v\pm \hat{s})^mE_m^{\underline{a}}$.

A bit more surprising might look the absence of
$e^{\beta(\xi)}=  \frac {\sqrt{2}} {\sqrt{(v-\hat{s})^{++}}}$ as this describes the $SO(1,1)$ transformations relating the  physical moving frame vectors $(v\pm \hat{s})^{\underline{a}}= (v\pm \hat{s})^mE_m^{\underline{a}}$ with constant frame \eqref{fu--fu++fuI}. This fact deserves more thought, and the structure and properties of the residual (super)symmetry \eqref{res-v-xi++}--\eqref{varTh-q===} deserve further study.

\section{On 'usual' tensionless super-$p$-branes and  their gauge fixing }

The  'usual' tensionless  super-p-branes, those which do not carry worldvolume gauge fields and often called {\it null-$p$-branes}, do exist as a classical model for a BPS object {\it in any number of spacetime dimension  $D$ and for any $p$}. This is because they are counterparts of the massless superparticles which are known to exist in any spacetime dimension   \cite{Bergshoeff:1996tu,Green:1999by}.
For definiteness, we will continue our discussion for {\it null--$p$--brane in 11D target superspace}. For our purposes it is convenient to use the spinor moving frame formulation of such 11D null-p-brane with the action
\be\label{Snull-p=}
S_{null-p}= \int d^{p+1}\xi\, \rho^{\# \, m}
E_m^{\underline{a}} \, u^=_{\underline{a}}\; .
\ee
It is a straightforward generalization of 11D null--superstring action ($p=1$) which was discussed in its relation with twistor and ambitwistor superstrings in \cite{Bandos:2006af} and \cite{Bandos:2014lja}.
The D=4 counterpart of this action had been proposed and studied in \cite{Bandos:1991my,Bandos:1993ma}, and a related formulation in the worldvolume superspace approach to null-superstring was the subject of \cite{Bandos:1993pa}.

In the action \eqref{Snull-p=} $u^=_{\underline{a}}(\xi) $ is a light-like vector field  which  is convenient to consider as a part of  moving frame described by $SO(1,10)$ valued matrix
\be\label{UinSO1-10} \left(\frac 1 2 \left(u^\#_{\underline{a}}+u^=_{\underline{a}}\right), u^I_{\underline{a}}, \frac 1 2 \left(u^\#_{\underline{a}}+u^=_{\underline{a}}\right)\right)\;\in\; SO(1,10)\; . \ee
However, in distinction  to the case of tensionless M5-brane, where $ u^=_{\underline{a}}$ is constructed from the flux and derivatives of PST scalar \eqref{u--=v-hs} (see \eqref{vsu=in}), in the case of usual null-p-branes it, as well as other moving frame vectors and spinor moving frame variables related to these by  Eqs. \eqref{u==v-v-=11D} --\eqref{uIts=v+v-=11D=} of Appendix \ref{app=smovingframe},  are independent constrained auxiliary fields \footnote{In the case of massless superparticle, $p=0$,  they can be related to the light-like momentum
\cite{Delduc:1991ir,Galperin:1991gk,Galperin:1992pz,Bandos:2007wm}. Actually, this statement applies also to null-p-branes with $p\geq 1$.}.
 $E_m^{\underline{a}}$ in \eqref{Snull-p=} is the pull--back of the bosonic supervielbein of the supergravity superspace (below we consider this to be flat)  to the $d=(p+1)$ dimensional  worldvolume and
$\rho^{\# \, m}(\xi)$ is an auxiliary field which transforms as worldvolume vector density under 6d diffeomorphisms, $m=0,1,...,p$.

\subsection{Equations  of motion and degeneracy of the induced metric }

The variation of the action with respect to the vector density $\rho^{\# \, m}(\xi)$ gives the equation
\be\label{Eu--=0}
E_m^{\underline{a}} \, u^=_{\underline{a}}=0\; ,
\ee
which implies, in particular, that the action \eqref{Snull-p=} vanishes on the mass shell. The equation for the fermionic coordinate functions can be simplified, when using the equations for $\delta X^{\underline{a}}$ discussed below, to
$ \rho^{\# \, m} \partial_m\Theta \Gamma^{\underline{a}} u_{\underline{a}}^==0$. Using constraints \eqref{u==v-v-=11D} this can be further simplified to
\be
\rho^{\# \, m} \partial_m\Theta^{\underline{\alpha}} v_{\underline{\alpha}q}^{\; -}=0\; .
\ee

The derivation or elaboration of all other equations requires either to vary with respect to constrained moving frame or spinor moving frame fields, or to calculate their derivatives. The simplest way to deal with these issues is to  use the expression for these fields in terms of a constant vector frame and constant spinor frame from Eqs. \eqref{u--=KuI=11}--\eqref{v-1+=fv-1+} of Appendix \ref{app=movingframe}. These are  similar to the expressions used in the case of tensionless M5--brane.

The relevant independent variations are then $\delta K^{= I}$,
which produces the equation
\be\label{rEfuI+=0}
\rho^{\# \, m} E_m^{\underline{a}}\, \left(\mathfrak{u}_{\underline{a}}^I+\frac 1 2 \, K^{=I}\mathfrak{u}_{\underline{a}}^\#\right)=0\; ,
\ee
while
$\delta \beta$ and $\delta {\cal O}_{qp}$ in variations of
\eqref{u--=fu--+} and  \eqref{v-1-=fv-1-} are parameters of $SO(1,1)$ and $SO(9)$ gauge symmetries of the action  \eqref{Snull-p=} (see sec. \ref{null-p=gaugeSymm}). The same applies to $\delta K^{\# I}$ in the variations of  \eqref{u++=Kui=11} and \eqref{uI=K--ui} which is the  parameter of the so-called ${\bb K}_9$ gauge symmetry which reflects the absence of moving frame vectors ${u}_{\underline{a}}^I$ and $u_{\underline{a}}^\#$ in the action \eqref{Snull-p=} (see sec. \ref{null-p=gaugeSymm} and e.g. \cite{Bandos:2007wm} for more detail). Setting this latter parameter to zero, we easily find that \eqref{rEfuI+=0} can be written in the form
\be\label{rEuI=0}
\rho^{\# \, m} E_m^{\underline{a}}u_{\underline{a}}^I=0\; .
\ee
This, together with \eqref{Eu--=0}, immediately implies that the induced metric
of the standard null-p-brane, $g_{mn}= E_m^{\underline{a}} \eta_{\underline{a}\underline{b}}E_n^{\underline{b}}$, is degenerate. Indeed, using \eqref{UinSO1-10} we can write it in the form
\be\label{gmn=dec} g_{mn}= E_m^{\underline{a}} \eta_{\underline{a}\underline{b}}E_n^{\underline{b}} =
 E_{(m}^{\underline{a}} u^{=}_{\underline{a}} E_{n)}^{\underline{b}} u_{\underline{b}}^\#  -  E_{m}^{\underline{a}} u^{I}_{\underline{a}} E_{n}^{\underline{b}} u_{\underline{b}}^I
\ee
in which it becomes transparent the existence of the on-shell null vector which is given by $\rho^{\# m}$
\be\label{rmgmn=0}
\rho^{\# m}g_{mn}=0\; .
\ee
This is tantamount to saying that this induced metric is degenerate on the mass shell.
Actually, this degeneracy can be shown by
using Eq. \eqref{Eu--=0} in  \eqref{gmn=dec} to write the on-shell induced metric as $ g_{mn}= E_m^{\underline{a}} \eta_{\underline{a}\underline{b}}E_n^{\underline{b}} =
 E_{(m}^{\underline{a}} u^{=}_{\underline{a}} E_{n)}^{\underline{b}} u_{\underline{b}}^\#  -  E_{m}^{\underline{a}} u^{I}_{\underline{a}} E_{n}^{\underline{b}} u_{\underline{b}}^I$. However, we have preferred to discuss  Eq. \eqref{rmgmn=0}
as this type of conditions of degeneracy naturally appears also in the ILST formulation of null-string \cite{Isberg:1993av}.

The equation of  motion for the bosonic coordinate function reads $\partial_m (\rho^{\# m} u^{=}_{\underline{a}} )=0$. Using \eqref{u--=fu--+} one can easily find that this implies two independent equation, one singlet and one nanoplet of SO(9):
\bea
\label{dre-2b=0}
\partial_m (e^{-2\beta}\rho^{\# m} )=0\; , \qquad
\\
\label{drK=0}\partial_m (\rho^{\# m} K^{=I})=0\; . \qquad
\eea

\subsection{Symmetries }

\subsubsection{Rigid symmetries }
The action \eqref{Snull-p=} is manifestly invariant and 11D superPoincar\'e
symmetry, including spacetime supersymmetry. It is also easy to check that it is invariant under the scaling symmetry  which acts, besides the coordinate functions, also on the moving frame  and spinor moving frame variables
\bea\label{scaling}
X^{\underline{a}}  \mapsto   \Lambda^2 X^{\underline{a}}\; , \qquad  \Theta^{\underline{\alpha}}  \mapsto   \Lambda \Theta^{\underline{\alpha}}\; , \qquad   \\
 u^{=}_{\underline{a}} \mapsto  \Lambda^{-2}u^{=}_{\underline{a}} , \quad   u^{\#}_{\underline{a}} \mapsto  \Lambda^{+2}u^{\#}_{\underline{a}} , \quad  u^{I}_{\underline{a}} \mapsto  u^{I}_{\underline{a}}\; ,  \qquad \\
 \rho^{\# m} \mapsto \rho^{\# m} \; . \qquad
\eea
For future reference, let us write explicitly the complete set of the rigid symmetry transformations  of the action \eqref{Snull-p=} {\it acting on the coordinate functions},
which is to say the above scaling (with $\Lambda =e^{2\lambda}$) and the 11D super-Poincar\'e symmetry
\bea
\label{vX=null-p}
\delta X^{\underline{a}} &=& a^{\underline{a}} + i \Theta\Gamma^{\underline{a}}\epsilon +  X^{\underline{b}}
L_{\underline{b}}{}^{\underline{a}}+2\lambda X^{\underline{a}} \; ,  \qquad  \\ \label{vTheta=null-p}
\delta \Theta^{\underline{\alpha}} &=& \epsilon^{\underline{\alpha}}  + \frac 1 4\, L^{\underline{a}\underline{b}} \, (\Theta\Gamma_{\underline{a}\underline{b}})^{\underline{\alpha}}
 +  \lambda \Theta^{\underline{\alpha}}  \; . \qquad   \eea

Like in the case of tensionless M5-brane, the complete conformal symmetry appears only in purely bosonic limit of the action \eqref{Snull-p=}, but is broken when   $\Theta^{\underline{\alpha}}$ is nonvanishing. This looks natural when one takes into account the fact that a spacetime superconformal symmetry is not known in 'usual' 11D superspace.  Indeed, in dimensions where that is known, the commutator of special conformal and supersymmetry generators produces conformal supersymmetry thus reproducing the full superconformal supergroup.

\subsubsection{Gauge symmetries and gauge fixing}
\label{null-p=gaugeSymm}

As $u_{\underline{a}}^I$ moving frame vectors do not enter the action, the  SO(9) rotation mixing these among themselves is a (trivial) manifest symmetry of the moving frame action \eqref{Snull-p=}. Moreover, as neither $u_{\underline{a}}^\#$ enters, also the so--called ${\bb K}_9$ symmetry mixing these with  $u_{\underline{a}}^I$
is also the gauge symmetry (see \cite{Bandos:2007wm} for more details and further references; in particular the action of gauge symmetries on spinor moving frame variables are presented there).

Let us stress that this  is different from the case of tensionless M5--brane where $u_{\underline{a}}^\#=(v+\hat{s})_{\underline{a}}$ is constructed from gauge field strength and derivative of PST scalar
(like $u_{\underline{a}}^==(v-\hat{s})_{\underline{a}}$
is) and hence ${\bb K}_9$ gauge symmetry is not present.
The same applies (due to the same reason) to the $SO(1,1)$ gauge symmetry of the standard null-p-brane action \eqref{Snull-p=} which acts by rescaling of $u_{\underline{a}}^=$ and opposite  rescaling of both $u_{\underline{a}}^\#$ and  $\rho^{\# m}$ \footnote{This last coincidence is reflected by the same double sign symbol
${}^\#\equiv {}^{++}$ of these two.},
\bea\label{SO1-1}
u_{\underline{a}}^= \mapsto e{}^{-2\tilde{\lambda}(\xi)} u_{\underline{a}}^=, \qquad
u_{\underline{a}}^\# \mapsto e{}^{+2\tilde{\lambda}(\xi)} u_{\underline{a}}^\# , \qquad \nonumber \\ \rho^{\# m}  \mapsto e{}^{+2\tilde{\lambda}(\xi)} \rho^{\# m} .\qquad 
\eea

Of course, using this gauge symmetry one can redefine the rigid spacetime scaling symmetry transformations \eqref{scaling} in such a way that these do not act on moving frame but do rescale $\rho^{\# m}$. Clearly, this cannot be done in the case of tensionless M5.

To see the form of the $\kappa$--symmetry leaving invariant the null-p-brane action \eqref{Snull-p=}, it is convenient to calculate the arbitrary fermionic variation of this which is defined by
\be
\delta_fX^{\underline{a}}=i \delta_f\Theta\Gamma^{\underline{a}}\Theta\; , \qquad \delta_f u_{\underline{a}}^==0\; , \qquad    \delta_f \rho^{m\#}=0 \; \qquad
\ee
with arbitrary
$\delta_f\Theta^{\underline{\alpha}}$.
One easily finds
\be
\delta_f S_{null-p}= -2i \int d\Theta \Gamma^{\underline{a}}\delta\Theta
u_{\underline{a}}^= = -4i \int d\Theta^{\underline{\alpha}}  v_{\underline{\alpha}q}^{\; -}  \, \delta_f\Theta^{\underline{\beta}}  v_{\underline{\beta}q}^{\; -}   \; , \qquad
\ee
where, to arrive at the final form, we have used constraints
\eqref{u==v-v-=11D} ({\it cf.} \eqref{v-s=v-v-} and \eqref{v-s=v-Gv-}).
Using the constraints \eqref{v+v+=0} one finds that  this vanishes for $\delta_f\Theta^{\underline{\alpha}} =  v_{q}^{-\underline{\alpha}} \kappa^{+ {q}}$ with an arbitrary fermionic function $\kappa^{+ {q}}(\xi)$.
Hence, taking into account Eqs. \eqref{v+v+=0}, we find that  the supersymmetry and $\kappa$--symmetry transformations of the fermionic coordinate function are given by Eq.
\eqref{vTh=ep+kqvq},
\be\label{vTh=ep+kqvq=}
\delta\Theta^{\underline{\alpha}} = \epsilon^{\underline{\alpha}} + v_q^{-\underline{\alpha}} \kappa^{+ {q}} \; .
\ee

At this point it becomes evident that the gauge fixing of the $\kappa$-symmetry and diffeomorphism symmetry of a null--$p$--branes of the standard type (which do not carry any worldvolume fields except for coordinate functions on their worldvolume), can be performed in the same manner as we did for tensionless M5--brane. The gauge fixed conditions for the coordinate functions can be chosen in the form of Eq. \eqref{Th+q=0} and \eqref{X++=xi++}--\eqref{Xi=xii} with $\check{m}=1,..., (p-1)$ and $\check{i}=1,..., (p-1)$.

The Goldstone fields are components of the fermionic coordinate function s which do not vanish  and bosonic coordinate functions which are not identified with worldvolume coordinates after the complete light-cone gauge fixing. These are sixteen fermionic  $\Theta^{-q}(\xi^m)=\Theta^{\underline{\alpha}} \mathfrak{v}^{-q}_{\underline{\alpha}}$, the same as  in the case of tensionless M5, but depending on $(p+1)$ worldvolume coordinates, $\xi^m=(\xi^{++},\xi^{--},\xi^{\check{i}})$ with  $\check{i}=1,..., (p-1)$,  and $(10-p)$ bosonic $X^{\check{I}}(\xi^m)$ enumerated by
$\check{I}=(p+1),...,10$.

The rigid symmetries preserving this light-cone gauge is described by exactly the same equations as in the case of tensionless M5--brane, but with the suitable range of the values of the indices described above.

  \bigskip

\section{Conclusions and discussion}

In this paper we have solved the problem of the gauge fixing of local fermionic $\kappa$--symmetry of the tensionless M5--brane the action of which was obtained in our recent \cite{Bandos:2026pdg}. To  this end we have used the spinor moving frame method  \cite{Bandos:1992np,Bandos:2007wm,Bandos:2014lja}  which can be adapted to the case of tensionless M5 brane in a quite nontrivial manner.

To clarify the difference between  tensionless M5-brane and standard null--$p$--branes which do not carry worldvolume gauge fields, we have also discussed spinor moving frame formulation of these latter. We have shown that the gauge fixed conditions for coordinate functions and residual symmetries preserving these conditions  are essentially the same as in the case of tensionless M5--brane and pointing out the differences.
In the case of  standard null-p--brane actions considered here, the  moving frame and spinor moving frame fields are auxiliary variables which can be associated with the momentum conjugate to the bosonic coordinate function. In the case of the tensionless M5--brane their meaning is different.

A peculiar feature of the tensionless limit of M5--brane \cite{Bandos:2026pdg} is that it requires to keep the worldvolume  gauge field strength nonvanishing and, moreover, nondegenerate. This latter implies that the antisymmetric second rank tensor
$\tilde{H}^{mn}$ constructed from  dual to the field strength $H_3$ of $b_2$ as it is shown in \eqref{tH=}, should have maximal number $2$ of nonvanishing eigenvalues. This is tantamount to saying that the composite 6--vector
$s_m$ defined in Eq. \eqref{sm:=} should be nonvanishing and thus allowing to construct  the spacelike unit vector $\hat{s}{}^{m}= \frac {s^m} {\sqrt{-s^ns_n}}$ which is necessary to define the $\kappa$-symmetry projector of the tensionless M5-brane in Eqs. \eqref{Gkappa-T0=}.

To construct the spinor moving frame cuitable for the light-cone  gauge fixing in the tensionless M5--brane model, we first lifted the unit worldvolume 6--vectors $\hat{s}_m$, constructed with the use of gauge field $b_2$, Eq. \eqref{sm:=}, and $v_m$, constructed from the derivative of PST scalar, Eq. \eqref{v:=}, to the unit 11-vectors $\hat{s}^{\underline{a}}=\hat{s}^mE_m^{\underline{a}}$ and $v^{\underline{a}}=v^mE_m^{\underline{a}}$, Eqs. \eqref{hsa:=}. Then we complete the set of these orthogonal unit vectors till the complete 11D  moving frame \eqref{vsu=in} by introducing nine spacelike unit vectors $u_{\underline{a}}^I$ orthogonal to that two and among themselves. The spinor moving frame  \eqref{VinSpin11} is defined  as a kind of square root of the vectors of such moving frame, namely
by the constraints
 \eqref{v-s=v-v-}--\eqref{v+s=v+Gv+}
and
\eqref{uIts=v+v-=11D}. Particularly, the constraints \eqref{v-s=v-v-}--\eqref{v+s=v+Gv+}
imply that spinor moving frame variables $v_{\underline{\alpha}q}^{\; \mp}$, which form the Spin(1,10) valued moving frame matrix, are a kind of square roots of light-like vectors given by the sum and the difference of $\hat{s}^{\underline{a}}$ and $v^{\underline{a}}$.

The use of spinor moving frame variables constructed in this way allowed us to fix the $\kappa$--symmetry of the tensionless M5--brane action by (a counterpart of the standard) light-cone gauge condition  \eqref{Th+q=0}. We have used these together with a constant spinor frame, related to some constant 'background field' configuration of the tensionless M5--brane, so that such gauge is fixed by $\kappa$--symmetry in a kind of neighborhood of this constant configuration defined by the conditions \eqref{det=v-fv}.

The light-cone gauge fixing of the worldvolume diffeomorphism symmetry in  \eqref{X++=xi++}-- \eqref{Xi=xii} is done with the use of the constant vector frame related to (constructed from) the above constant spinor frame. In this light-cone gauge the equations of motion for the bosonic and fermionic coordinate functions become  formally similar to the  superstring or null--superstring equations but with a non-holonomic derivative $\nabla_{++}$ of Eq. \eqref{nabla++==} which, however, depends  nonlinearly on the derivatives of the coordinate functions.

We have also presented the residual diffeomoprphism symmetry transformations  \eqref{res-v-xi++} --\eqref{res-v-xii}
preserving the light-cone  gauge of Eqs.
\eqref{Th+q=0} and \eqref{X++=xi++}-- \eqref{Xi=xii}. The constant parameters of these are given by the parameters of the target space Poincar\'e supergroup which are split with the use of constant vector frame and associated spinor frame, and of the scaling symmetry.
The residual diffeomorphism symmetry transformations of the worldvolume coordinates \eqref{X++=xi++}--\eqref{Xi=xii} in this case involve the Goldstone fields of the tensionless M5--brane. This phenomenon was observed in \cite{Bandos:2026pdg} for the bosonic limit of the tensionless M5--brane.

In its turn, the transformations of the Goldstone fields of the gauge fixed tensionless M5--brane involve the contributions proportional to the worldvolume coordinates, first of all through the residual diffeomorphisms in \eqref{res-v-XchI} and  \eqref{varTh-q===}, but in the case of bosonic Goldstone field
also through the gauge fixed version of the target space Lorentz transformations (see the first line in  Eq.  \eqref{res-v-XchI=}).

As was noticed in \cite{Bandos:2026pdg}, in the tensionless limit the worldvolume 2-form gauge field $b_2$ of M5--brane becomes inert  under its local fermionic $\kappa$--symmetry. As a worldvolume field without bulk spacetime symmetry indices, it is also  invariant  under spacetime supersymmetry. At first glance this might suggest a problem related to the balance of degrees of freedom usually required by a (linearly realized) supersymmetry. This problem, however, does not appear in the case of tensionless branes as their $\kappa$--symmetry is similar to the one of massless superparticles which, as it is well known, exists in any number of dimensions in spite of a dis-balance which occurs e.g. in D=11 (see \cite{Bergshoeff:1996tu,Green:1999by}).

\begin{widetext}
After gauge fixing the spacetime supersymmetry transformations do act on the worldvolume gauge field through the residual diffeomorphism transformations,
\be\label{vbmn=}
{\underline\delta} b_{mn}=b'_{mn}(\xi)-b_{mn}(\xi)=-\delta\xi^lH_{lmn}+2\partial_{[m}\Big(b_{n]l}\delta\xi^l\Big)+ 2 \partial_{[m} \alpha_{n]}\,
\ee
with $\delta\xi^m=(\delta \xi^{++}, \delta \xi^{--}, \delta \xi^{\check{i}}) $ given in \eqref{res-v-xi++} -- \eqref{res-v-xi++} and 1--form gauge symmetry parameter $\alpha_{n}(\xi)$.
This can be used to remove the second term so that the induced supersymmetry transformation of the worldvolume gauge field reads
\bea\label{vbmn=susy}
{\underline\delta}_{susy} b_{mn}=-\delta_{susy} \xi^lH_{lmn}
= -2i \Theta^{-q(\mathfrak{v})}\left(\epsilon^{-q(\mathfrak{v})} + \frac 1 2 \nabla_{++} X^{\check{I}}\gamma^{\check{I}}_{qp} \epsilon^{-p(\mathfrak{v})}\right)H_{--\, mn}+2i \Theta^{-q(\mathfrak{v})}\gamma^{\check{i}}_{qp}\epsilon^{+p(\mathfrak{v})}H_{\check{i}mn}
\, . \qquad
\eea
\end{widetext}

Although the number of supersymmetries which preserve the gauge fixed version of the tensionless M5-brane action in flat target superspace is the same as in maximal superconformal group in 6 dimensions, the bosonic symmetries are not manifestly that of superconformal group.

In general, the structure of symmetries preserving the light-cone gauge fixed on the coordinate functions of tensionless M5-brane is quite complicated, which makes difficult to handle the supergroup they form. However, it is not difficult to observe that all these originate in flat superspace super-Poincar\'e symmetry completed by symmetry under Weyl rescaling of spacetime coordinate functions.
Probably the study of tensionless M5-brane in $AdS_{7(4)}\times S^{4(7)} $ or $AdS_2\times Y^9 $ superspace allows  to see a bigger symmetry. Notice that even in the simplest version of the latter case, maximally supersymmetric
$AdS_2\times S^2\times T^7$ solution of 11D supergravity, the superspace description  is still to be elaborated.

\bigskip

\subsection*{Acknowledgements}
The author is thankful to Dima Sorokin  and Kurt Lechner for the collaboration at early stages of this project and useful discussions.
This work  was supported in part by the Spanish  MCI, AEI and FEDER (ERDF EU) grant  PID2024-155685NB-C21``Gravity, Supergravity and Superstrings'' (GRASS), as well as by the Basque Government Grant ”Gravitation, Cosmology and Fundamental Physics” IT1977-26.

\newpage
\begin{widetext}
\appendix

\centerline{{\large\bf Appendices}}

\section{Some properties of D=11 gamma matrices}
\label{app=11DGamma}

In our mostly minus metric convention, $\eta_{ab}= diag(+,-,...,-)$, the 32$\times 32$ 11D Gamma matrices $(\Gamma^{{a}})_{\underline\alpha}{}^{\underline{\beta}}$ obeying the
Clifford algebra, $\{\Gamma^a \; , \Gamma^b\}= 2 \eta^{ab}$, are imaginary. The charge conjugation matrix
$C_{\underline{\alpha}\underline{\beta}}=- C_{\underline{\beta}\underline{\alpha}}$ and its inverse
$C^{\underline{\alpha}\underline{\beta}}=- C^{\underline{\beta}\underline{\alpha}}$ are imaginary as well.
We find more convenient to use  real and symmetric matrices
\begin{eqnarray}
\Gamma^{{a}}_{\underline{\alpha}\underline{\beta}} =
(\Gamma^{{a}})_{\underline{\alpha}}{}^{\underline{\gamma}}C_{\underline{\gamma}\underline{\beta}}=
\Gamma^{{a}}_{\underline{\beta}\underline{\alpha}}\; , \qquad \\ 
\tilde{\Gamma}_{{a}}{}^{\underline{\alpha}\underline{\beta}} = C^{\underline{\alpha}\underline{\gamma}}
(\Gamma_{{a}})_{\underline{\gamma}}{}^{\underline{\beta}}=
\tilde{\Gamma}_{{a}}{}^{\underline{\beta}\underline{\alpha}}\; , \qquad
\end{eqnarray}
which obey
\begin{eqnarray}
\Gamma^{{a}}\tilde{\Gamma}^{{b}}+  \Gamma^{{b}}\tilde{\Gamma}^{{a}}= 2\eta^{ab}\, . \qquad
\end{eqnarray}
Then for Majorana spinors  $\epsilon\Gamma^a\psi = \epsilon^\alpha  \Gamma^a_{\alpha\beta}\psi ^\beta =
\epsilon_\gamma\tilde{\Gamma}^{a\gamma\delta}\psi_\delta= \epsilon^\alpha  \Gamma^a{}_{\alpha}{}^{\beta}\psi
_\beta $.

The (anti)symmetrization of the indices is defined with overall weight unity, so that  \begin{eqnarray}
\tilde{\Gamma}^{{a}_1...a_{2p+1}}= \tilde{\Gamma}^{[{a}_1} ... \tilde{\Gamma}^{a_{2p+1} ]}= \frac {1}{ (2p+1)!}\left( \tilde{\Gamma}^{{a}_1} ... \tilde{\Gamma}^{a_{2p +1}}
\mp  permutations \right)\, ,  \qquad \\
\tilde{\Gamma}^{{a}_1...a_{2p}}= \tilde{\Gamma}^{[{a}_1} ... {\Gamma}{}^{a_{2p} ]}=\frac {1}{(2p)!}\left( \tilde{\Gamma}^{{a}_1} ... \Gamma^{a_{2p} }
\mp  permutations \right)\, .  \qquad
\end{eqnarray}

We choose
\begin{eqnarray}
&& \Gamma_{0\ldots 9\; {\rm 10}} = -i \, I \; , \qquad \tilde{ \Gamma}_{0\ldots 9\; {\rm 10}} = +i \, I \; , \qquad \epsilon_{0\ldots 9\; {\rm 10}}=1= \epsilon^{0\ldots 9 \; 10}\;
\end{eqnarray}
which implies
\begin{eqnarray}
 \label{G=e*GA}
 \Gamma_{a_1\ldots a_n} &=&
\frac {i (-1)^{\frac {n(n-1)}{2}}}{(11-n)!} \epsilon_{a_1\ldots a_nb_1\ldots b_{11-n}}
 \Gamma^{b_1\ldots b_{11-n}} \; .
\end{eqnarray}

\section{More details on spinor moving frame }

\label{app=movingframe}

\subsection{Complete set of constraints on spinor moving frame variables}

\label{app=smovingframe}

The complete list of constraints on  spinor moving frame variables, forming the Spin(1,10) valued matrix
\eqref{VinSpin11}, \be\label{VinSpin11=app}
\left(v_{\underline{\alpha} q}{}^+ , v_{\underline{\alpha} q}{}^- \right)\; \in\; Spin(1,10)\; ,
\ee
and moving frame variables forming Lorentz group valued matrix \be\label{UinSO1-10=App} \left(\frac 1 2 \left(u^\#_{\underline{a}}+u^=_{\underline{a}}\right), u^I_{\underline{a}}, \frac 1 2 \left(u^\#_{\underline{a}}+u^=_{\underline{a}}\right)\right)\;\in\; SO(1,10)\;  \ee
and, thus, obeying and \eqref{u--u--=0}--\eqref{epu1-u9=},
are
(see e.g. \cite{Bandos:2007wm} and refs. therein)
\begin{eqnarray}\label{u==v-v-=11D}
 &
  v^-_{{q}} \tilde{\Gamma}_{{\underline{\mu}}}v^-_{{p}}=u_{\underline{\mu}}^= \delta_{{q}{p}}   \; , & \qquad  u_{\underline{\mu}}^= \Gamma^{\underline{\mu}}_{\underline{\alpha}\underline{\beta}}= 2v_{\underline{\alpha} q}{}^- v_{\underline{\beta} q}{}^-  \; ,  \qquad \\ \nonumber \\
\label{v+v+=u++}
&  \;  v_{{q}}^+ \tilde{\Gamma}_{{\underline{\mu}}} v_{{p}}^+ =u_{{\underline{\mu}}}^{\# } \delta_{{q}{p}}\; , & \qquad u_{ {\underline{\mu}}}^{\# } {\Gamma}^{ {\underline{\mu}}}_{ {{\underline{\alpha}}} {{\underline{\beta}}}} =2 v_{{{\underline{\alpha}}}{q}}{}^{+}v_{{{\underline{\beta}}}{q}}{}^{+}  \; ,  \qquad \\ \nonumber  \\
 \label{uIs=v+v-=11D=}
& v_{{q}}^- \tilde {\Gamma}_{{\underline{\mu}}} v_{{p}}^+=u_{ {\underline{\mu}}}^{I} \gamma^I_{q{p}}\; , &\qquad
 u_{{\underline{\mu}}}^{I} {\Gamma}^{{\underline{\mu}}}_{{\underline{\alpha}}{\underline{\beta}}} =2 v_{( {{\underline{\alpha}}}|{q} }{}^- \gamma^I_{qp}v_{|{{\underline{\beta}}}){p}}{}^{+} \; , \quad    \qquad \\ \nonumber  \\
 \label{uIts=v+v-=11D=}
& v_{{q}}^- {\Gamma}_{{\underline{\mu}}} v_{{p}}^+=-u_{ {\underline{\mu}}}^{I} \gamma^I_{q{p}}\; , &\qquad
 u_{{\underline{\mu}}}^{I} \tilde{\Gamma}^{{\underline{\mu}}\underline{\alpha}\underline{\beta}} =-2 v_q^{-(\underline{\alpha}} \gamma^I_{qp}v_p^{+\underline{\beta})} \; . \quad  \end{eqnarray}
 In
 \eqref{uIs=v+v-=11D=} and  \eqref{uIts=v+v-=11D=}
 $\gamma^I_{qp}=\gamma^I_{p{q}}$ are SO(9) Dirac matrices with properties listed in
\eqref{gIgJ+=I}.

The above equations describe also the relations of moving frame with the elements of the inverse spinor moving frame matrix,
\be
\left(\begin{matrix} v^{-\underline{\alpha}}_q \cr v^{-\underline{\alpha}}_q \end{matrix}\right)\; \in\; Spin(1,10)\;
\ee
which obey \eqref{v+v+=0}--\eqref{I=vv+vv} and are expressed through the spinor moving frame variables \eqref{VinSpin11} by Eqs. \eqref{V-1=CV-A}.

In the case of tensionless M5-brane, we construct the two light-like moving frame vectors from derivatives of the PST scalar (in $v_m$ of \eqref{v:=}) and gauge field strength (in $\hat{s}_m$ of \eqref{sm:=}, \eqref{hsm:=}) according to  \eqref{u--=v-hs},
\be\label{u--==v-hs=App}  u_{\underline{a}}^\# = (v+\hat{s})_{\underline{a}}\; ,  \qquad
 u_{\underline{a}}^= = (v-\hat{s})_{\underline{a}}\; .  \qquad
\ee
\subsection{Relation between two moving frames and two  spinor moving frames }
\label{app:v-fv}

The complete form of the relations between two moving frames is described by the following equation obtained in  \cite{Bandos:2017eof} (see Eqs. (7.19)-(7.31) of this paper):
 \begin{eqnarray}\label{u--=KuI=11}
 u^=_{\underline{a}} &=&e^{-2\beta}\left(\mathfrak{u}^=_{\underline{a}} + \frac 1 4 (\vec{K}^{=} )^2 \mathfrak{u}^\#_{\underline{a}}  + K^{=I}\mathfrak{u}^I_{\underline{a}} \right)
 \; , \qquad \\
 \label{u++=Kui=11}
 u^\#_{\underline{a}}&=& e^{+2\beta}\left(\mathfrak{u}^\#_{\underline{a}} \left(1+ \frac 1 2 K^{=I}{K}{}^{\# I}+ \frac 1 {16} (\vec{K}{}^{=})^2 (\vec{K}{}^{\#})^2\right)    + \frac {1}{4} \mathfrak{u}^=_{\underline{a}} (\vec{K}{}^{\#})^2 +  \mathfrak{u}^I_{\underline{a}} \left( {K}{}^{\#  I}+ \frac 1 4 K^{=I} (\vec{K}{}^{\#})^2 \right)   \right)
, \; \qquad \\
 \label{uI=K--ui}
u^I_{\underline{a}} &=& \left(\mathfrak{u}^K_{\underline{a}}
 \left(\delta^{KJ} + \frac 1 2 K^{=K}{K}{}^{\#  J} \right)
+  \frac 1 2 \mathfrak{u}^\#_{\underline{a}} \left( K^{= J }+ \frac 1 4 {K}{}^{\#  J} (\vec{K}{}^{=})^2 \right)  +  \frac 1 2 \mathfrak{u}^=_{\underline{a}}  \tilde{K}{}^{\#  J}
  \right){\cal O}^{JI} \; . \qquad
\end{eqnarray}
In Eqs. \eqref{u--=KuI=11}--\eqref{uI=K--ui} ${K}{}^{=J}$ and ${K}{}^{\#  J} $ parametrize coset $SO(1,10)/[SO(1,1)\times SO(9)]$, $\beta $ is the parameter of SO(1,1) subgroup of SO(1,10) and
${\cal O}^{JI}={\cal O}^{-1\,IJ}$ is the SO(9) valued matrix.

In this paper we use these relation to relate the moving frame with a constant (and, hance, not moving) frame. In the case of tensionless M5-brane, the light--like vectors of moving frame are expressed in terms of worldvolume gauge field strength and derivative of PST scalar by Eqs. \eqref{u--==v-hs=App}.

The expression for the spinor moving frame fields in terms of constant spinor harmonics are
\begin{eqnarray}\label{v-=fv+Kfv}
v_{\underline{\alpha} q}^{\; -} &=& e^{-\beta} {\cal O}_{{q}{p}}  \left(
\mathfrak{v}_{\underline{\alpha}  p }^{\; -}+ \frac {1}{2} K^{=I}  \gamma^I_{pp'} \mathfrak{v}_{\underline{\alpha}  p }^{\; +}
\right) \; , \qquad  \\
\label{v+=fv+Kfv}
{v}_{\underline{\alpha}  q }^{\; +} &=& e^{\; \beta} {\cal O}_{{q}{p}}
\mathfrak{v}_{\underline{\alpha}  p }^{\; +} +  \frac{1}{ 2} e^{\,2\, \beta}\,  {\cal O}^{IJ} {K}^{\# J} \gamma^I_{pp'} \mathfrak{v}_{\underline{\alpha}  p' }^{\; -}
\qquad \\  &=& e^{\; \beta} {\cal O}_{{q}{p}}\left(
\mathfrak{v}_{\underline{\alpha}  p }^{\; +} +\frac 1 4 {K}^{\# J} {K}^{= I} (\gamma^J\gamma^I)_{pp'} \mathfrak{v}_{\underline{\alpha}  p' }^{\; +} + \frac {1}{2}  {K}^{\# I} \gamma^I_{pp'} \mathfrak{v}_{\underline{\alpha}  p' }^{\; -}\right)
  \; . \qquad
\end{eqnarray}
These expressions involve $Spin(9)$ matrices related to SO(9)  matrix used above by
\begin{eqnarray}\label{OIJ==OGO}
 {\cal O}_{qr} {\cal O}_{{p}{s}} \gamma^{I}_{rs}= \gamma^{J}_{qp} {\cal O}^{JI}\; .
   \qquad
\end{eqnarray}
Furthermore, as $Spin(9)\subset SO(16)$ these matrices  also obey
\begin{eqnarray}\label{cOT==cO-1}
 {\cal O}_{qp} = {\cal O}^{-1}_{{p}{q}} \qquad\; .
   \qquad
\end{eqnarray}

The expression for the elements of the inverse spinor moving frame matrix read

\begin{eqnarray}\label{v-1-==fv-1-}
v_{q}^{-\underline{\alpha}} &=& e^{-\beta} {\cal O}_{{q}{p}} \left(
\mathfrak{v}_{{p} }^{ -\underline{\alpha}}- \frac {1}{2} K^{=I} \gamma^I_{p{p}'} \mathfrak{v}_{{p}' i}^{+\underline{\alpha}}
\right) \; , \qquad  \\
\label{v-1+=fv-1+}
v_{{q}}^{+\underline{\alpha}}&=& e^{\; \beta} {\cal O}_{{q}{p}}\left(\left( \delta_{pp'}+ \frac 1 4  {K}^{\# I}  {K}^{=J} (\gamma^I\gamma^J)_{pp'} \right)
\mathfrak{v}_{ {p}'i}^{+\underline{\alpha}} - \frac 1 2  {K}^{\# I}  \gamma^I_{pp'}
\mathfrak{v}_{ {p}'i}^{-\underline{\alpha}}\right)
  \; .\qquad
\end{eqnarray}

\bigskip

\section{Derivation of Eq. \eqref{vX=gaugeTh}}
\label{derivation=vXa}

After fixing the gauge \eqref{Th+q=0}, the  transformations  of bosonic spacetime symmetries and supersymmetry induce some $\kappa$--symmetry transformations so that the generic transformation  of the bosonic coordinate functions acquire the form of Eq.   \eqref{vX=gaugeTh}. This can be derived as follows
\bea
\label{vX=gaugeTh=deriv}
\delta X^{\underline{a}} &=& a^{\underline{a}} +  X^{\underline{b}}
L_{\underline{b}}{}^{\underline{a}}+\lambda X^{\underline{a}} + i \Theta\Gamma^{\underline{a}}\epsilon - i \Theta\Gamma^{\underline{a}}\kappa\qquad \nonumber \\ &=& a^{\underline{a}} +  X^{\underline{b}}
L_{\underline{b}}{}^{\underline{a}}+\lambda X^{\underline{a}} +i\Theta^{-q(\mathfrak{v})}
\left(
\mathfrak{u}^{\# \underline{a}}\epsilon^{-q(\mathfrak{v})}  + \mathfrak{u}^{I \underline{a}}\gamma^I_{qp}\epsilon^{+p(\mathfrak{v})}
\right)
- \qquad \nonumber \\ && \qquad - i\Theta^{-q(\mathfrak{v})}\gamma^I_{qr}\left(\mathfrak{u}^{I \underline{a}}- \frac 1 2 K^{=I}\mathfrak{u}^{\# \underline{a}}\right)\, e^{-\beta} {\cal O}_{rp}\kappa^{+p}=\qquad \nonumber \\  &=& a^{\underline{a}} +  X^{\underline{b}}
L_{\underline{b}}{}^{\underline{a}}+\lambda X^{\underline{a}} +i\Theta^{-q(\mathfrak{v})}
\left(
\mathfrak{u}^{\# \underline{a}}\epsilon^{-q(\mathfrak{v})}  + \mathfrak{u}^{I \underline{a}}\gamma^I_{qp}\epsilon^{+p(\mathfrak{v})}
\right)
+ \qquad \nonumber \\ && \qquad + i\Theta^{-q(\mathfrak{v})}\gamma^I_{qp}\left(\mathfrak{u}^{I \underline{a}}- \frac 1 2 K^{=I}\mathfrak{u}^{\# \underline{a}}\right)\left(\epsilon^{+p(\mathfrak{v})}+ \frac 1 2 L^{\# I} \Theta^{-s(\mathfrak{v})} \gamma^I_{sp} \right)=\qquad \nonumber  \\
 &=& a^{\underline{a}} +  X^{\underline{b}}
L_{\underline{b}}{}^{\underline{a}}+\lambda X^{\underline{a}} +i\Theta^{-q(\mathfrak{v})}
\left(
\mathfrak{u}^{\# \underline{a}}\epsilon^{-q(\mathfrak{v})}  + \mathfrak{u}^{I \underline{a}}\gamma^I_{qp}\epsilon^{+p(\mathfrak{v})}
\right)
+ \qquad \nonumber \\ && \qquad + i\left(\Theta^{-q(\mathfrak{v})}\gamma^I_{qp}\epsilon^{+p(\mathfrak{v})}+ \frac 1 2L^{\# J} \Theta^{-q(\mathfrak{v})}\gamma^{IJ}_{qp} \Theta^{-p(\mathfrak{v})}  \right) \left(\mathfrak{u}^{I \underline{a}}- \frac 1 2 K^{=I}\mathfrak{u}^{\# \underline{a}}\right)
\; .  \qquad
\eea

\bigskip

\section{On off-shell description of the gauge fixed tensionless M5 }

\label{app:off-shell}

The off-shell form of the gauge fixed theory is  useful, in particular, if one searches for the gauge fixed form of the action, besides that of the equations of motion. In this case we have to use
the complete Eq.
\eqref{v-hs=gauge=--} instead of its simplified form \eqref{v-hs=gauge=--on}. Eqs.  \eqref{v-hs=gauge=}-- \eqref{v-hs=gauge=cI},
imply that
\be\label{vd-sd=cD}
(v-\hat{s})^m \partial_m = (v-\hat{s})^{++} {\cal D}_{++}
\ee
with
 \bea
  {\cal D}_{++}&=& {\nabla}_{++} + \, \frac {{\nabla}_{++}\Theta^{-q} \, \Theta^{-q}} {1-2i \partial_{--}\Theta^{-q} \, \Theta^{-q} }\, \partial_{--} \, \qquad \nonumber \\
  &=& \partial_{++}+ \frac 1 4 ({K}^{=\check{I}})^2\partial_{--} + \, \frac {\partial_{++}\Theta^{-q} \, \Theta^{-q}
  +\frac 1 4 ({K}^{=\check{I}})^2\partial_{--}\Theta^{-q} \, \Theta^{-q}
  } {1-2i \partial_{--}\Theta^{-p} \, \Theta^{-p} }\, \partial_{--} \, . \qquad
 \eea

In the first line of this equation we use  ${\nabla}_{++}$ defined in terms of ${K}^{=\check{I}}$ in Eq. \eqref{nabla=K}. Off the mass--shell \eqref{vd-sd=cD}
and \eqref{(v-s)dXI=-KI} imply

\bea \label{KI=-cDXI} K^{=\check{I}}=-2 {\cal D}_{++}  X^{\check{I}}\; , \qquad {\check{I}}=5,6,7,8,9\;  \eea
so that applying ${\cal D}_{++}$ to $ X^{\check{I}}$
we arrive at a kind of algebraic equation for this derivative,
\be\label{cD++XI=}
{\cal D}_{++} X^{\check{I}} = \partial_{++} X^{\check{I}} +  \left(({\cal D}_{++} X^{\check{J}})^2+   \frac {
\partial_{++}\Theta^{-q} \, \Theta^{-q}
  + ({\cal D}_{++} X^{\check{K}})^2 \,\partial_{--}\Theta^{-q} \, \Theta^{-q} } {1-2i\partial_{--}\Theta^{-q} \, \Theta^{-q} }
  \right)  \partial_{--} X^{\check{I}}\, .
\ee
Clearly, for a 'chiral' bosonic Goldstone field obeying   $\partial_{--} X^{\check{I}}=0$, ${\cal D}_{++} X^{\check{I}}$ reduces to $\partial_{++} X^{\check{I}} $
\be
{\cal D}_{++} X^{\check{I}}\vert{_{\partial_{--} X^{\check{J}}=0}} = \partial_{++} X^{\check{I}}\: .
\ee

In the generic case, the square of
\eqref{cD++XI=} gives us a quadratic equation for
$Y=({\cal D}_{++} X^{\check{I}})^2$,
\be\label{Y2+}
(D_{--}X^{\check{J}})^2\, Y^2- (1-2D_{++}X^{\check{J}}D_{--}X^{\check{J}}) Y + (D_{++}X^{\check{J}})^2=0\; ,
\ee
where we have introduced notation
\bea\label{D--=}
D_{--} &=& (1+\Xi)\,\partial_{--} \; , \qquad \\ \label{D++XI=} && \Xi\, =  \, \frac {\partial_{--}\Theta^{-q} \, \Theta^{-q}} {1-2i \partial_{--}\Theta^{-p} \, \Theta^{-p} }\; , \qquad
\\ \label{D++=}
D_{++} &=& \partial_{++}+
\, \frac {\partial_{++}\Theta^{-q} \, \Theta^{-q}} {1-2i \partial_{--}\Theta^{-p} \, \Theta^{-p} } \partial_{--}\; .
\eea
In the assumption of $(\partial_{--}X^{\check{J}})^2\not=0$,
Eq. \eqref{Y2+} is solved by
\bea\label{Y+-=}
Y_{\pm} &=& \frac 1 {2 (1+\Xi)^2 (\partial_{--}X^{\check{K}})^2}\left( 1-2 (1+\Xi)D_{++}X^{\check{J}}\partial_{--}X^{\check{J}}  \right. \qquad \nonumber \\ && \left. \pm \sqrt{\left( 1-2 (1+\Xi)D_{++}X^{\check{J}}\partial_{--}X^{\check{J}} \right)^2- (1+\Xi)^2 (D_{++}X^{\check{J}})^2(\partial_{--}X^{\check{I}}})^2\right)\; . \qquad
\eea
Formally we can consider both
$({\cal D}_{++}X^{\check{J}})^2=Y_+$ and $({\cal D}_{++}X^{\check{J}})^2=Y_-$ cases, but actually, only the second is consistent.

Indeed, let us consider the case
of $\partial_{++}X^{\check{J}}=0=\partial_{++}\Theta^{-q(\mathfrak{v})}$ in which
$D_{++}X^{\check{J}}=0$ so that $Y_-$ vanishes while $Y_+= \frac 1 {(1+\Xi)^2 (\partial_{--}X^{\check{J}})^2}$. Although this latter looks consistent in our assumption of $ (\partial_{--}X^{\check{J}})^2\not=0$ used to obtain \eqref{Y+-=},
it  does not look as a reasonable expression for $(D_{++}X^{\check{J}})^2$ in the whole theory, as this clearly has the trivial solution
$D_{++}X^{\check{J}}=0=\partial_{--}X^{\check{J}}$ for which the expression for $Y_+$ becomes inconsistent with its definition.

Finally, the off--shell expression for the characteristic derivative of the coordinate functions is, in the above notation,
\be
{\cal D}_{++}= D_{++} + Y_-(1+\Xi)\partial_{--}\; .
\ee

\bigskip

\end{widetext}

\providecommand{\href}[2]{#2}\begingroup\raggedright\endgroup

\end{document}